\documentclass[12pt,preprint]{aastex}
\usepackage{color}

\newcommand{\etal}{et\,al.}
\newcommand{\halpha}{H$\alpha$}
\newcommand{\lsim}{\raise0.3ex\hbox{$<$}\kern-0.75em{\lower0.65ex\hbox{$\sim$}}}
\newcommand{\gsim}{\raise0.3ex\hbox{$>$}\kern-0.75em{\lower0.65ex\hbox{$\sim$}}}

\newcommand{\msun}{M$_{\odot}$}

\newcommand{\HI}{H~{\sc i}}
\newcommand{\kms}{km\,s$^{-1}$}

\newcommand{\hr}{{\it h}$_{\rm r}$}
\newcommand{\hthree}{{\it h}$_{\rm 3.6}$}
\newcommand{\hfour}{{\it h}$_{\rm 4.5}$}
\newcommand{\hhi}{{\it h}$_{\rm HI}$}
\newcommand{\rhi}{R$_{\rm HI}$}
\newcommand{\adbs}{ADBS\,113845+2008}
\begin{document}
\slugcomment{Accepted for publication in the Astronomical Journal}
\title{Exploring the Interstellar Media of Optically Compact Dwarf Galaxies}
   




\author{Hans P. Most, John M. Cannon}
\affil{Department of Physics \& Astronomy, Macalester College, 1600 Grand 
Avenue, Saint Paul, MN 55105}
\email{hmost@macalester.edu, jcannon@macalester.edu}

\author{John J. Salzer}
\affil{Department of Astronomy, Indiana University, 727 East Third Street, 
Bloomington, IN 47405}
\email{slaz@astro.indiana.edu}

\author{Jessica L. Rosenberg}
\affil{School of Physics, Astronomy, and Computational Science, George Mason University,
Fairfax, VA 22030}
\email{jrosenb4@gmu.edu}

\author{Eric Engstrom}
\affil{Department of Physics \& Astronomy, Macalester College, 1600 Grand 
Avenue, Saint Paul, MN 55105}

\author{Palmer Fliss}
\affil{Department of Physics \& Astronomy, Macalester College, 1600 Grand 
Avenue, Saint Paul, MN 55105}
\affil{Department of Earth Sciences, University of Minnesota, Minneapolis, MN, 55455}

\begin{abstract}

We present new {\it Very Large Array} \HI\ spectral line, archival
{\it Sloan Digital Sky Survey}, and archival {\it Spitzer Space
  Telescope} imaging of eight star-forming blue compact dwarf galaxies
that were selected to be optically compact (optical radii \textless 1
kpc). These systems have faint blue absolute magnitudes (M$_{\rm B}
\gsim -$17), ongoing star formation (based on emission-line selection
by the \halpha\ or [\ion{O}{3}] lines), and are nearby (mean velocity
$=$ 3315 km\,s$^{-1}$ $\simeq$ 45 Mpc). One galaxy in the sample, ADBS
113845+2008, is found to have an \HI\ halo that extends 58 r-band
scale lengths from its stellar body.  In contrast, the rest of the
sample galaxies have \HI\ radii to optical-scale-length ratios ranging
from 9.3 to 26. The size of the \HI\ disk in the ``giant disk'' dwarf
galaxy ADBS 113845+2008 appears to be unusual as compared to
similarly compact stellar populations.

\end{abstract}						


\section{Introduction}
\label{S1}

Blue compact dwarf galaxies (hereafter, BCDs) have emerged as a
particularly intriguing subset of dwarf galaxies that have been
studied in the local universe (z$\lsim$0.05).  Since the early studies
and characterizations by \citet{thuan81} and \citet{thuan83}, many
investigations have classified BCDs as systems that harbor spatially
and temporally concentrated ongoing star formation in a low-mass
galaxy potential well.  While definitions in the literature vary, BCDs
are instrinsically faint systems (M$_{\rm B} \gsim-$17) that exhibit
ongoing star formation over the bulk of an otherwise physically
compact stellar population.

Detailed studies of individual, nearby BCDs have provided important
insights into their physical properties and into the mechanisms that
initiate and govern their star formation.  \citet{vanzee98} found that
five compact dwarf galaxies that are currently experiencing strong
bursts of star formation exhibit central \HI\ surface densities that
are significantly higher than those of dwarf irregular (dIrr)
galaxies.  Similarly, some well-studied BCDs harbor very high column
density neutral hydrogen gas in the regions of active star formation
(e.g., N$_{\rm HI}$ $\simeq$ 2.4\,$\times$\,10$^{22}$ cm$^{-2}$ in
I\,Zw\,18; see {Brown \etal\ 2002}\nocite{brown02} and {van~Zee
  \etal\ 2006}\nocite{vanzee06}).  Other recent studies have verified
that many nearby and well-resolved systems have high neutral gas
surface densities in the regions of ongoing star formation (see, e.g.,
{Thuan \etal\ 2004}\nocite{thuan04}, {Elson
  \etal\ 2010}\nocite{elson10}, and references therein).

The physical mechanism(s) that create this central concentration of
neutral gas remain(s) ambiguous.  One commonly proposed solution is
that the current burst events in BCDs are the result of a
gravitational interaction with a nearby companion
\citep[e.g.,][]{ostlin01}.  While there are examples of apparently
interaction-driven burst events (e.g., the SBS\,1129$+$576 system
studied by {Ekta \etal\ 2006}\nocite{ekta06}), there are also systems
that have been shown to be quite isolated (see, for example, {van~Zee
  \etal\ 1998}\nocite{vanzee98}, {Bureau
  \etal\ 1999}\nocite{bureau99}, and further discussion below). While
undiscovered, very low-mass companions may reside near some of these
isolated systems (consider, for example, the discovery of an
interaction with a low surface brightness companion to the Magellanic
irregular NGC\,4449 by {Mart{\'{\i}}nez-Delgado
  \etal\ 2012}\nocite{martinezdelgado12}), the gravitational
interaction trigger does not appear to be universal in BCD galaxies.
The origin of the concentrated star formation in at least some of the
isolated BCDs appears to require a different initiation mechanism.

While the properties of isolated BCDs have been studied in some detail
(e.g., the studies of NGC\,2915 by {Meurer
  \etal\ 1994}\nocite{meurer94}, {1996}\nocite{meurer96}), a
particularly intriguing system was discovered in the {\it Arecibo Dual
  Beam Survey} (ADBS) by \citet{rosenberg00,rosenberg02}: \adbs.  This
galaxy was subsequently cataloged as a UV-excess source in the 2nd
KISO survey ({Miyauchi-Isobe \& Maehara, 2000}\nocite{miyauchi00}).
Its extremely compact nature was first recognized during {\it WIYN}
0.9-meter\footnote{The {\it WIYN} 0.9-meter telescope is operated
  jointly by a consortium that includes Indiana University, San
  Francisco State University, the University of Wisconsin-Madison, the
  University of Wisconsin-Stevens Point, the University of
  Wisconsin-Whitewater, Austin Peay State University, Haverford
  College, the Rochester Institute of Technology, and the Wisconsin
  Space Grant Consortium.} broadband imaging of selected ADBS galaxies.
The optical component was found to be extremely compact (B-band
exponential scale length $=$0.57 kpc); \halpha\ narrowband imaging
confirmed a high ongoing star formation rate (0.06 \msun\,yr$^{-1}$)
within this optically compact stellar component.

Subsequent {\it Very Large Array}\footnote{The National Radio
  Astronomy Observatory is a facility of the National Science
  Foundation operated under cooperative agreement by Associated
  Universities, Inc.} ({\it VLA}) \HI\ spectral line imaging of the
source revealed that the neutral gas disk was extremely extended
compared to the compact optical structure.  As discussed in detail in
\citet{cannon09}, \HI\ gas was detected to a maximum radial distance
of $\sim$25 kpc at the 10$^{19}$ cm$^{-2}$ level; it thus possesses
one of the largest \HI-to-optical size ratios of any known dwarf
galaxy.  In contrast to some detailed \HI\ studies of other BCD
systems that have found complex velocity structure
\citep[e.g.,][]{vanzee98,simpson11}, \adbs\ is undergoing well-ordered
rotation throughout most of the neutral gas disk.

Perhaps the most mysterious quality of \adbs\ is that its neutral gas
morphology is dominated by a broken ring structure that is roughly 15
kpc in diameter.  This ring contains the highest column density
neutral gas in the galaxy and is not coincident with the optical
component.  Rather, the star-forming component resides in a relative
\HI\ under-density.  \citet{cannon09} interpret this result to be a
signature of an extremely inefficient mode of star formation in a very
isolated galaxy.  We hereafter refer to this striking collection of
physical properties as the ``giant disk'' phenomenon in dwarf
galaxies.

The remarkable characteristics of \adbs\ are the origin of the present
investigation.  Stated most simply, we seek to understand if this
``giant disk'' scenario is common or rare in star-forming dwarf
galaxies that have stellar components that are as strongly centrally
concentrated as the one in \adbs.  To this end, we present new
optical, infrared, and \HI\ imaging of a sample of extremely optically
compact dwarf galaxies.  The new observations presented here confirm
the rarity of the ``giant disk'' found in the \adbs\ system.

\section{Observations and Data Reduction}
\label{S2}
\subsection{Survey Selection Criteria}
\label{S2.1}

The sample of optically compact galaxies studied in this paper was
produced using specific selection criteria designed to identify
star-forming galaxies with compact stellar components; \adbs\ and
IC\,2271 were already known to have such characteristics from previous
imaging.  Starting with the full {\it KPNO International Spectroscopic
  Survey} ({\it KISS}, an objective-prism survey for extragalactic
emission-line objects; Salzer \etal\ 2000\nocite{salzer00},
2001\nocite{salzer01}, 2002\nocite{salzer02}) database, we searched
for emission-line objects that were spectroscopically-confirmed.  To
maximize physical resolution an upper velocity constraint was set at
V$_{\rm sys}$ $<$ 5,000 \kms.  Photometric parameters were obtained
for these objects from the {\it SDSS} using the {\it
  SkyTools}\footnote{See
  http://cas.sdss.org/astro/en/tools/crossid/upload.asp} interface.
Objects with optical radii less than 3\arcsec\ were selected; these
were then individually inspected to ensure a compact optical
structure. Finally, we searched for archival single-dish
\HI\ observations of these remaining systems in order to determine
their suitability for {\it VLA} observing.  This resulted in a sample
of six galaxies selected from the {\it KISS} catalogs.  We stress that
this sample is meant to be representative of the compact dwarf class,
but it is not complete in the sense of being a volume-limited sample
of such sources.

All of these galaxies share the following characteristics: 1) optical
compactness (optical radii at 50\% peak brightness of $\lsim$1 kpc);
2) faint blue absolute magnitudes (M$_B \gsim -$17); 3) ongoing star
formation (based on emission-line selection by the \halpha\ or
[\ion{O}{3}] lines); and 4) proximity for spatially resolved work with
the {\it VLA} C-array (15\arcsec\ beam $\lsim$5 kpc for all systems;
mean velocity $=$ 3315 km\,s$^{-1}$ $\simeq$ 45 Mpc).  Note that in
the work of \citet{cannon09}, \adbs\ was selected using these same
criteria.  It is important to emphasize that the six {\it KISS}
galaxies in the present study (see Table~\ref{t1} for salient
properties) were selected based on their optical properties alone; the
\HI\ flux integrals were used only to determine observational
feasibility.

\subsection{{\it VLA} \HI\ Spectral Line Observations}
\label{S2.2}

\HI\ spectral line imaging of each of the sample galaxies was obtained
with the {\it VLA} in the C configuration via observing programs AC
841 and AC 919 (PI: Cannon).  The {\it VLA} correlator was used in a
standard single polarization mode, providing 64 or 128 channels over a
1.56 or 3.12 MHz bandwidth.  The velocity resolution (2.56, 5.15, or
10.3 km\,s$^{-1}$\,ch$^{-1}$) was selected based on the (single dish)
line width of each source.  Combinations of bandwidth and frequency
resolution were selected to guarantee sufficient line-free channels to
allow an accurate removal of the continuum emission.  Table~\ref{t2}
provides details about each observing session from which data are
presented in this work.

All reductions were performed using the Astronomical Image Processing
(AIPS) package.  Due to the array being in a ``hybrid'' state,
containing both {\it VLA} and {\it EVLA} receivers, standard reduction
procedures were followed with necessary modifications to account for
the ``aliasing'' effect that arose due to the digital-to-analog
conversion required to correlate {\it EVLA} signals using the old {\it
  VLA} correlator.  This effect was most egregious for narrow
bandwidth observations.  We thus followed the NRAO instructions of
flagging {\it EVLA}-{\it EVLA} baselines before and unflagging them
after the tasks BPASS and CALIB\footnote{See
  https://science.nrao.edu/facilities/vla/obsolete/aliasing for more
  details}. The resulting continuum subtraction was performed using a
first-order fit in UVLSF to channels where the bandpass was
demonstrably flat.  We found no measurable differences between this
calibration approach and one where we disregarded the {\it EVLA}
antennas completely and followed standard reduction procedures.  We
note that there were only 5 {\it EVLA} antennas in the array during
the C configuration observation of \adbs\ (which would contain the
most measurable aliasing effect, given the bandwidths listed in
Table~\ref{t2}).

Two data cubes were produced for each galaxy using the task IMAGR and
each one was cleaned to 2 times the rms noise (H\"ogbom
1974\nocite{hogbom74}; Clark 1980\nocite{clark80}).  First,
``low-resolution'' cubes were created using natural weighting (ROBUST
= 5 in the AIPS IMAGR task).  These cubes were convolved to a circular
beam size of 20\arcsec\ x 20\arcsec.  Second, ``high-resolution''
cubes were created using robust weighting (ROBUST = 0.5; see {Briggs
  1995}\nocite{briggs95}).  These cubes were convolved to a circular
beam size of 15\arcsec\ x 15\arcsec.

To create blanked cubes and moment maps, the naturally weighted cubes
were convolved to a beam size of 30\arcsec.  These cubes were blanked at
the 2 $\sigma$ level found in line-free channels.  The cubes were then
inspected by hand, and emission in three or more consecutive channels
was considered to be real.  These cubes were then used to blank both
the ``low-'' (20\arcsec) and ``high-resolution'' (15\arcsec)
cubes. Through this procedure, we assure that the same pixels
contribute to both the ``low-'' and the ``high-resolution''
cubes. Moment maps showing integrated \HI\ emission and velocity
structures were created using the techniques discussed in detail in
\citet{cannon09}.

As seen in Table~\ref{t3}, the total \HI\ flux integrals derived for
the sample galaxies are each lower than the {\it Arecibo Observatory}
single dish observations \citep{rosenberg00,rosenberg02,lee02}. The
cause of these differences is at least partly due to the missing short
{\it uv} spacings of the {\it VLA} compared to a single dish
measurement.  For example, the observed \HI\ flux integral of IC\,2271
of 1.22 Jy km\,s$^{-1}$ is significantly lower than the value found by
\citet{rosenberg00} using the {\it VLA} D array (4.7 Jy km\,s$^{-1}$).
Similarly, the total \HI\ flux integral derived for \adbs\ (1.58 Jy
km\,s$^{-1}$) is lower than the 2.58 Jy km\,s$^{-1}$ found by
\citet{cannon09} using combined C and D configuration data.  A
re-reduction of the D configuration data sets as part of the present
analysis confirms a higher flux integral using the more compact
configuration (S$_{\rm HI}$ = 2.12\,$\pm$\,0.21 Jy\,km\,s$^{-1}$ for
ADBS 113845+2008 and S$_{\rm HI}$ = 1.61\,$\pm$\,0.16 Jy\,km\,s$^{-1}$
for IC 2271, both with 55\arcsec\ circular beam sizes).  Since the
KISSR galaxies (which constitute the bulk of the present sample) have
only C configuration observations, we chose to intentionally omit the
D configuration data for IC\,2271 and \adbs\ from the present
analysis; this assures that all eight galaxies have similar thresholds
for surface brightness sensitivity at different spatial scales.

In an effort to recover as much flux from each of the target galaxies
as possible, we performed various tests with the {\it VLA} {\it uv}
datasets.  Long baselines were downweighted in IMAGR via
tapering in the {\it uv} plane.  Circular beam size cubes with
resolution elements of 15\arcsec, 30\arcsec, and 60\arcsec\ were
created for each galaxy using appropriate UVTAPER values.  The results 
of these tests are summarized in Table~\ref{t4}.  There are only modest 
gains in the recovered flux integrals for most sources.  KISSR\,278 
shows the most pronounced increase in S$_{\rm HI}$ when moving from 
15\arcsec\ to 30\arcsec\ resolution.  Note that IC\,2271 and \adbs, 
which have the largest single dish flux integrals (see Table~\ref{t3}),
show only modest increases in S$_{\rm HI}$ when downweighting long 
baselines.  This is strong evidence that the missing shortest spacing
baselines are important for the recovery of all of the \HI\ flux from these
systems.

We also experimented with calibrating our {\it uv} data using only
certain baselines in the array.  Using KISSR\,561 as a template, we
independently calibrated the {\it uv} data in five different ways: by
removing the {\it VLA}-{\it EVLA} baselines; by removing the {\it
  VLA}-{\it VLA} baselines; by retaining only the {\it EVLA}-{\it
  EVLA} baselines; by retaining only the {\it VLA}-{\it VLA}
baselines; and by the previously discussed method in which the {\it
  EVLA}-{\it EVLA} baselines were flagged and unflagged around
specific tasks. We find no significant differences in the galaxy's
morphology amongst these tests.  The last method (i.e., the flagging
and unflagging of {\it EVLA}-{\it EVLA} baselines) proved to recover
the most flux, and so was used for all datasets in the sample. 

As we discuss in detail in \S~\ref{S3.2} below, five of the eight
sample galaxies have nearby companions, some of which are very
\HI-bright.  A third possible explanation for the lower flux integrals
derived for some of the systems studied here is that these
high-surface-brightness companions introduce sidelobes, the imperfect
cleaning of which will prevent a full recovery of the flux integral
from the dim target galaxies.  In fact, this effect can lead to the
single dish flux measurements being erroneously high.  The system that
is most egregiously affected in this way is KISSR\,396; as we discuss
below, this source is a companion to the massive spiral NGC\,5375.  It
is possible that the single dish measurement of KISSR\,396 is an
over-estimate, since NGC\,5375 is in the first sidelobe of the
single-dish primary beam.  In an attempt to image this complex field
with high fidelity, we implemented cleaning algorithms using both
multi-faceted cleans around the detected galaxies and multi-scale
clean algorithms \citep{cornwell08}.  Neither approach produced
significant increases in the flux integral recovered from KISSR\,396.
We thus use the {\it VLA} S$_{\rm HI}$ values quoted in Table~\ref{t3}
throughout the remainder of this work.

\subsection{{\it SDSS} and {\it Spitzer} Observations}
\label{S2.3}

Optical and infrared imaging of the sample galaxies was obtained from
public archival databases to facilitate detailed comparisons with the
\HI\ imaging presented here.  This was necessary because the imaging
data obtained as part of {\it KISS} \citep{salzer00,salzer01} had an
image scale of 2.0\arcsec\,pixel$^{-1}$; none of the {\it KISS}
galaxies studied here are resolved in the survey data.  Optical r-band
images were obtained from the {\it SDSS} Data Release Eight
\citep{aihara11}.  3.6 $\mu$m and 4.5 $\mu$m {\it Spitzer} images of
the six KISSR galaxies were obtained from the {\it Spitzer Heritage
  Archive} ({\it Spitzer} imaging of IC\,2271 and \adbs\ is not
available).  No further treatments of the pipeline-processed archival
data were required.  These images are discussed in further detail in
the sections below.

Throughout this paper we will discuss the physical sizes of the
galaxies in this sample.  We thus draw specific attention to the
angular resolution of the optical and infrared imaging used in our
analysis.  The FWHM of the {\it Spitzer}/IRAC images is
1.66\arcsec\ at 3.6 $\mu$m and 1.72\arcsec\ at 4.5 $\mu$m\footnote{see
  the IRAC Instrument Handbook at http://ssc.spitzer.caltech.edu}.
Similarly, the median PSF FWHM of {\it SDSS} r-band images is quoted
as 1.3\arcsec\footnote{www.sdss.org}, and we adopt this value here;
using the individual images we analyze in this work, we measure 
empirical PSF FWHM values ranging from 0.94\arcsec\ (IC\,2271) to
1.22\arcsec\ (KISSR\,245).

\section{\HI\ and Stellar Distributions}
\label{S3}
\subsection{Target Galaxies}
\label{S3.1}

\begin{figure}
\epsscale{1.0}
\plotone{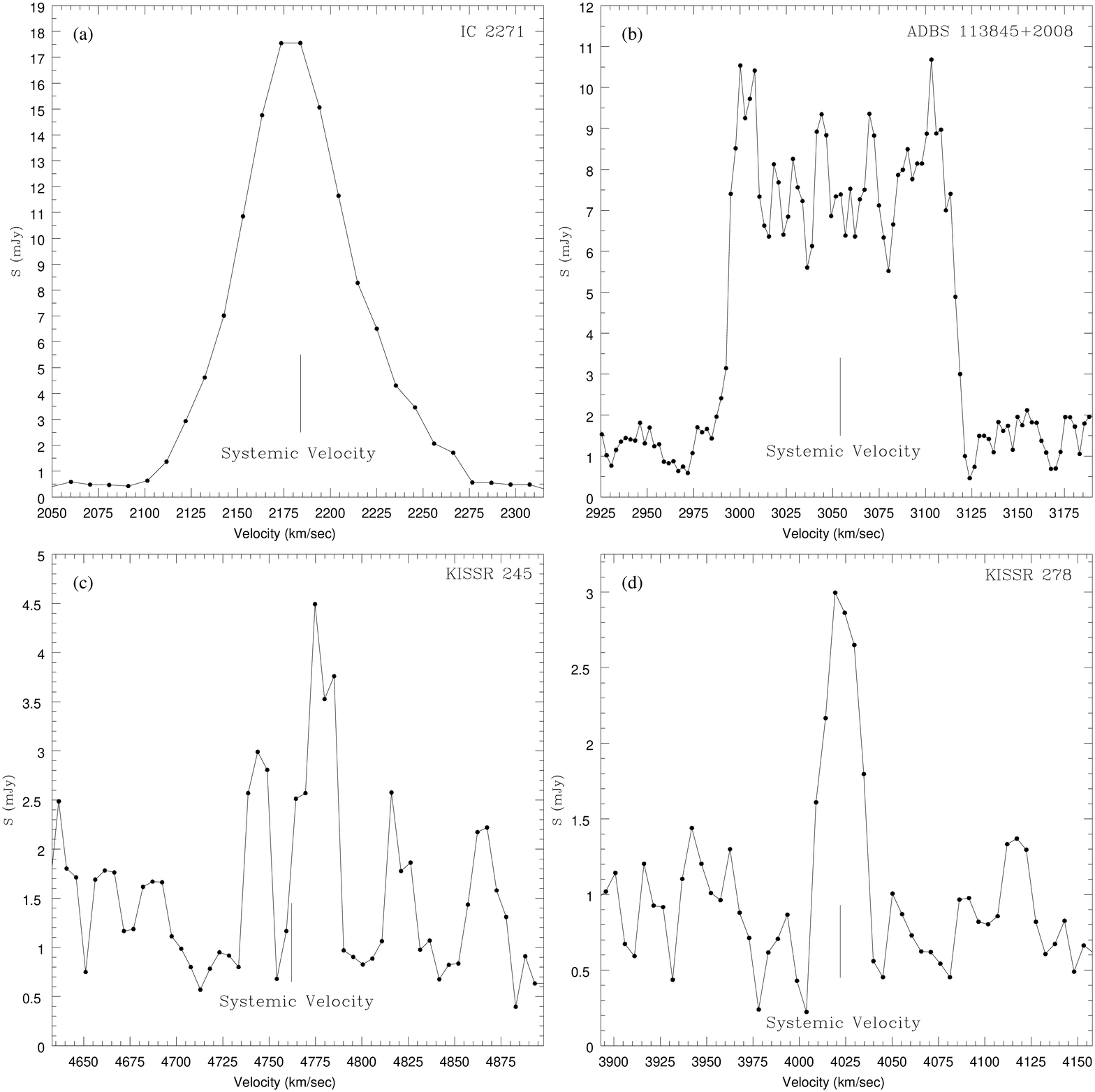}
\epsscale{1.0}
\caption{Global \HI\ profiles for IC\,2271 (a), \adbs\ (b), KISSR\,245
  (c), and KISSR\,278 (d), with the derived systemic velocities
  shown. These profiles were produced by determining the flux in each
  channel after the by-hand blanking process discussed in
  \S~\ref{S2.2}.  The flux in each channel was boxcar averaged with
  the flux from the two adjacent channels to create the profiles shown
  here.}
\label{figcap1}
\end{figure}

\begin{figure}
\epsscale{1.0}
\plotone{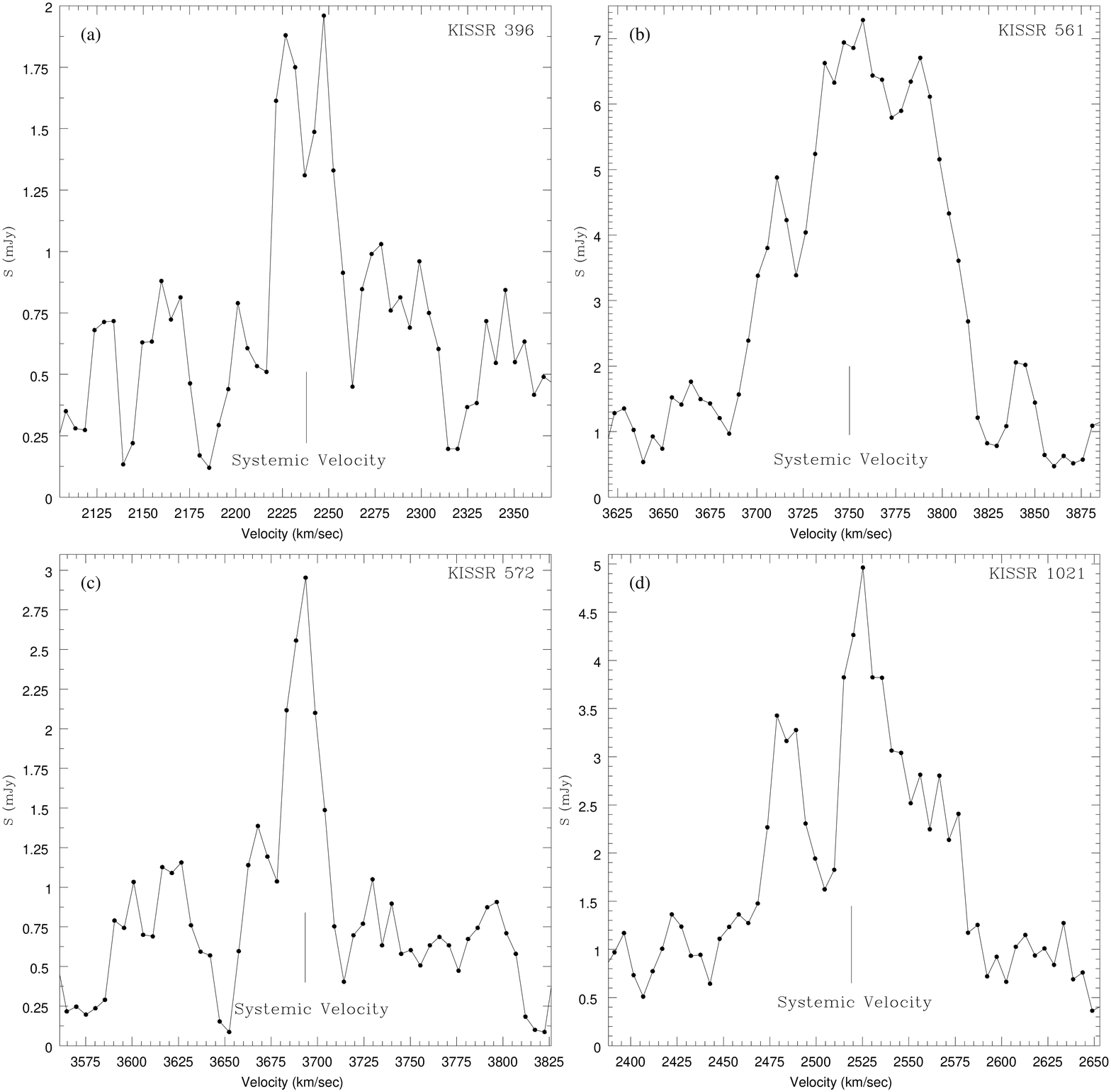}
\epsscale{1.0}
\caption{Same as Figure~\ref{figcap1}, for KISSR\,396 (a), KISSR\,561 (b), KISSR\,572 (c), and KISSR\,1021 (d).}
\label{figcap2}
\end{figure}

Using the blanked datacubes discussed in \S~\ref{S2.2}, real emission
in each channel was summed in order to create the global \HI\ profiles
shown in Figures~\ref{figcap1} and \ref{figcap2}.  It is immediately
evident that the neutral gas disks of the sample galaxies span a wide
range of physical characteristics.  Two systems show broad
\HI\ profiles ($\Delta$V $\gsim$100 km\,s$^{-1}$) that are indicative
of bulk rotation (e.g, \adbs, KISSR\,561).  Three systems show what
may be double-horned or broad profiles that are measured with poor
fidelity (e.g., KISSR\,245, KISSR\,396, KISSR\,1021).  Three systems
show simple single-peak profiles indicative of either low rotation
velocities or nearly face-on inclination angles (e.g, IC\,2271,
KISSR\,278, KISSR\,572).

The profiles shown in Figures~\ref{figcap1} and \ref{figcap2} should
be compared with the recovered flux integrals presented in
Table~\ref{t3} and discussed in \S~\ref{S2.2}.  The three systems with
the smallest recovered S$_{\rm HI}$ values in Table~\ref{t3}
(KISSR\,278, KISSR\,396, KISSR\,572) are those systems with the
narrowest \HI\ profiles.  Note specifically that KISSR\,278,
KISSR\,396, and KISSR\,572 are each only detected in 4, 7, and 3
consecutive channels of their datacubes, respectively.  It is
important to keep the low signal to noise ratio of some of these
galaxies in mind during the discussion that follows.

Single dish profiles of the six KISSR galaxies in
Figures~\ref{figcap1} and \ref{figcap2} were presented in
\citet{lee02}.  A comparison of these two sets of profiles shows
similar morphologies in some cases; the profiles of KISSR\,278,
KISSR\,396, and KISSR\,561 shown in Figures~\ref{figcap1} and
\ref{figcap2} are similar in shape as those shown in \citet{lee02},
although with slightly weaker peak flux densities in the
interferometric global profiles.  Conversely, the other three KISSR
systems (KISSR\,245, KISSR\,572, and KISSR\,1021) show significant
differences in the shapes of their profiles; the single-dish profiles
not only recover more flux (see detailed discussion in \S~\ref{S2.2}
above), but they also recover different morphologies (note, for
example, that the single dish profile of KISSR\,1021 shows a
pronounced double-horn morphology that is only poorly recovered in the
interferometric global profile).

\begin{figure}
\epsscale{1.0}
\plotone{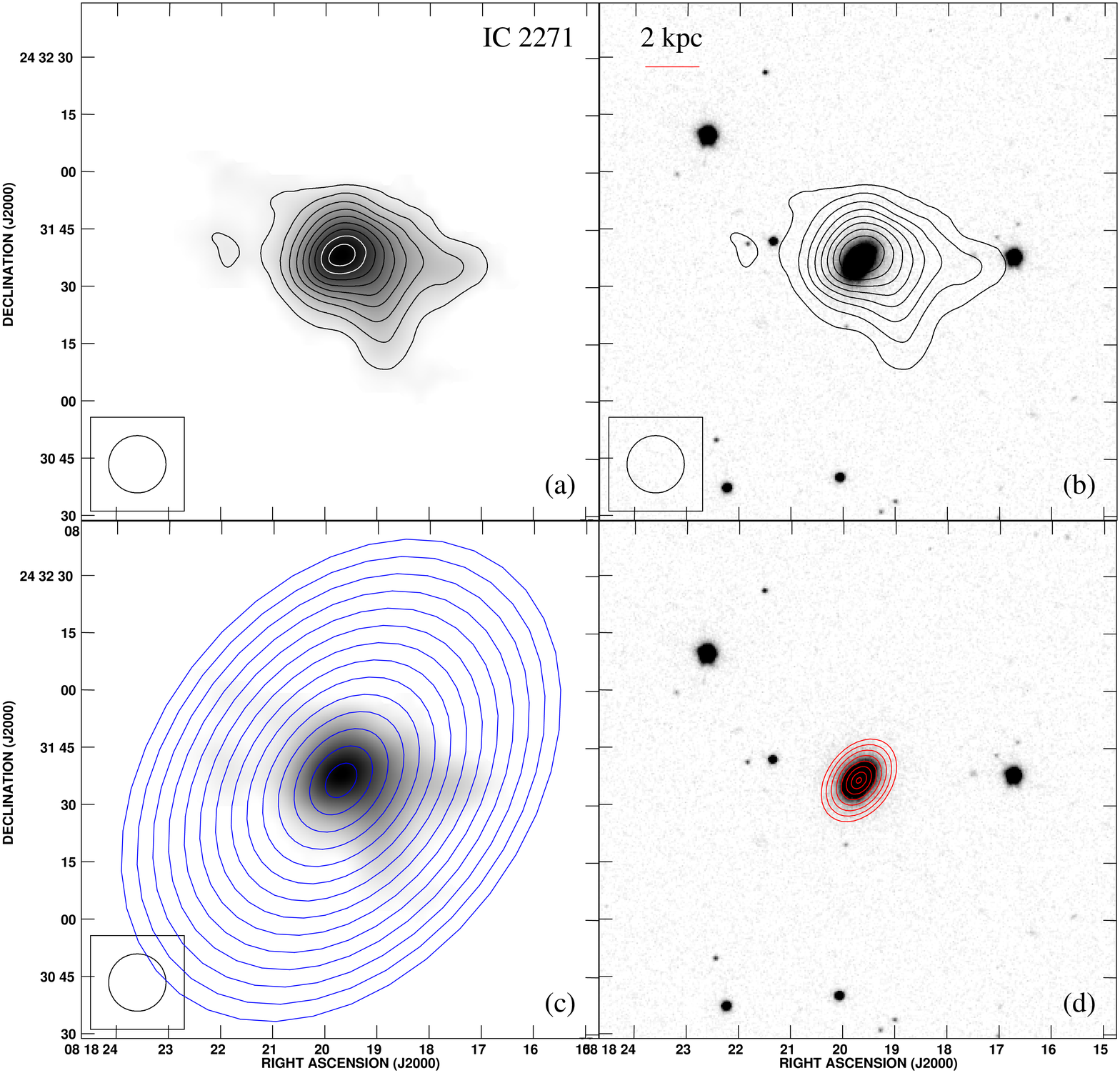}
\epsscale{1.0}
\caption{{\it VLA} and {\it SDSS} images of IC\,2271. Panels (a) and
  (c) show the \HI\ column density image at
  15\arcsec\ resolution. Panels (b) and (d) show the {\it SDSS} r-band
  image. The \HI\ column density contours in panels (a) and (b) are at
  the (2, 4, 6, 8, 10, 12, 14, 16, 18)\,$\times$\,10$^{20}$ cm$^{-2}$
  levels. The ellipses are those used to determine the \HI\ (c; see
  \S~\ref{S4.2}) and optical (d; see \S~\ref{S4.1}) radial
  profiles. The ellipses in panel (d) are shown as every fourth to be
  visible at this size.  The red bar in panel (b) denotes a physical
  length of 2 kpc at the adopted distance.}
\label{figcap3}
\end{figure}

\begin{figure}
\epsscale{1.0}
\plotone{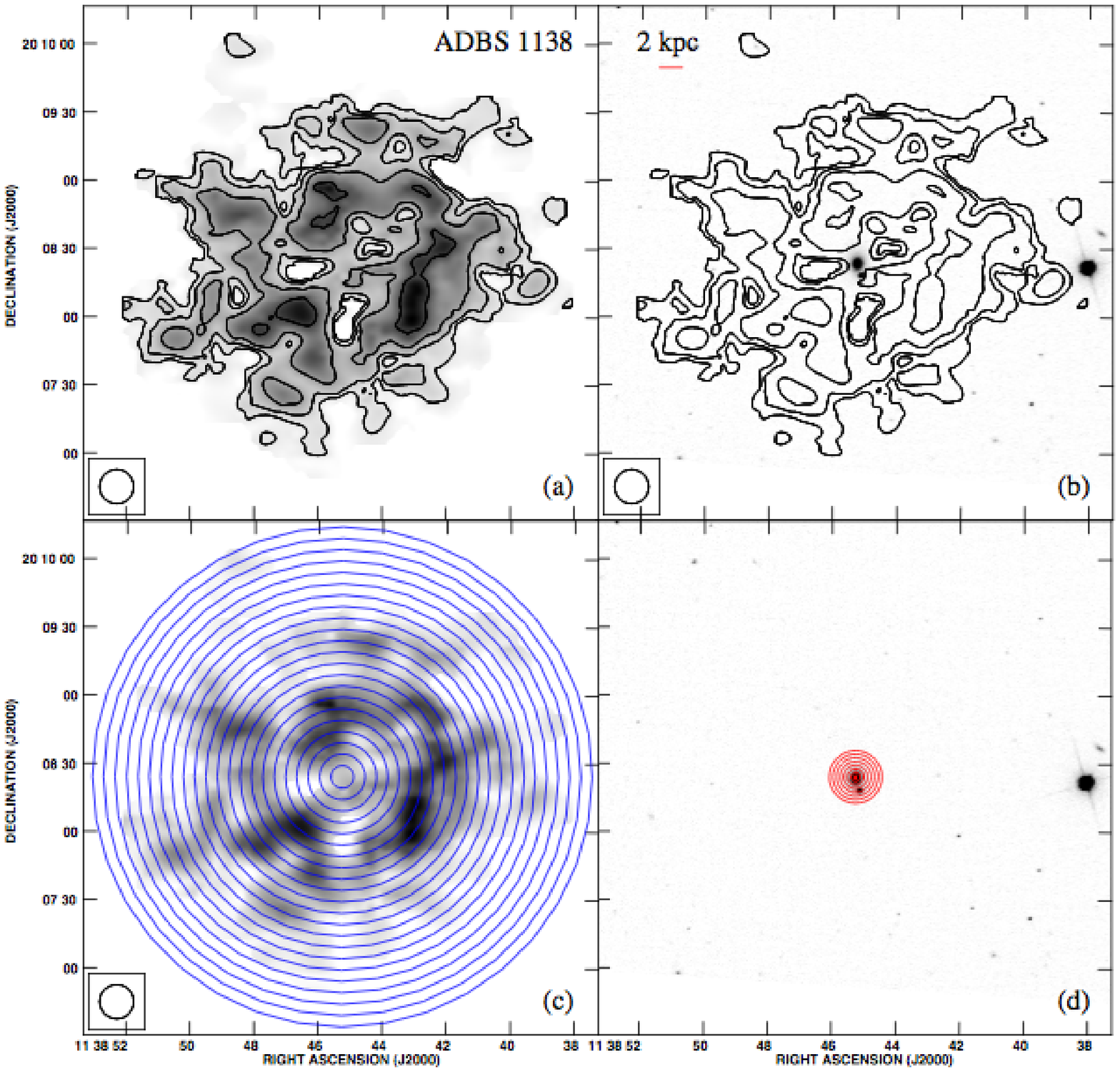}
\epsscale{1.0}
\caption{{\it VLA} and {\it SDSS} images of \adbs. Panels (a) and (c)
  show the \HI\ column density image at 15\arcsec\ resolution. Panels
  (b) and (d) show the {\it SDSS} r-band image. The \HI\ column
  density contours in panels (a) and (b) are at the (0.2, 0.4, 0.8,
  1.6)\,$\times$\,10$^{20}$ cm$^{-2}$ levels.  The ellipses are those
  used to determine the \HI\ (c; see \S~\ref{S4.2}) and optical (d;
  see \S~\ref{S4.1}) radial profiles The ellipses in panel (d) are
  shown as every fourth to be visible at this size.  Note the
  extremely optically compact nature of \adbs, especially when
  compared to the foreground star immediately to the southwest of the
  galaxy.  The red bar in panel (b) denotes a physical length of 2 kpc
  at the adopted distance.}
\label{figcap4}
\end{figure}

Figures~\ref{figcap3} through \ref{figcap10} compare the \HI, optical,
and infrared (available for six of the eight systems, excluding
IC\,2271 and \adbs) images of the eight sample galaxies.  Using the 2
kpc physical length denoted in panel (b) of each figure, it is
immediately evident that the high surface brightness stellar component
of each of these eight galaxies is extremely compact.  In fact, some
of these galaxies are only slightly more extended than foreground
stars found in the fields [note the close proximity of foreground
  stars to \adbs\ (Figure~\ref{figcap4}) and KISSR\,561
  (Figure~\ref{figcap8})]; we discuss the effects of resolution on our
results in more detail below.  While the \HI\ surface brightness
varies from one system to the next, each of these eight galaxies is
detected at high significance in both the {\it SDSS} and in the {\it
  Spitzer} images.

An examination of the \HI\ column density images and contours in
Figures~\ref{figcap3} through \ref{figcap10} reveals that all of the
galaxies in this sample, with the exception of \adbs, harbor centrally
concentrated neutral gas distributions at these spatial resolutions.
In most cases the optical and infrared surface brightness maxima are
spatially coincident with the highest detected column density region
(e.g., IC\,2271 in Figure~\ref{figcap3}, KISSR\,572 in
Figure~\ref{figcap9}); in some systems there is a small offset between
these two maxima, but most of these offsets are smaller than the
synthesized beam sizes and are thus not significant.  The most
significant discrepancy between the stellar and gaseous components is
found in \adbs: as discussed in detail in \citet{cannon09}, the
\HI\ morphology is dominated by a broken ring structure that surrounds
the optical body of the system.  The stellar component is many kpc
away from the highest column density neutral gas.

Similarly, an examination of the optical and infrared images of these
sources reveals that most members of the sample are extremely compact.
However, it is important to stress that each of the systems is
spatially resolved in the {\it SDSS} images, the {\it Spitzer} images,
or both; we discuss this in more detail in \S~\ref{S4.1} below.  A few
of the galaxies do have low surface brightness emission in the optical
and infrared that may be indicative of the compact burst being only
the highest surface brightness component of a larger stellar disk.
The most obvious systems with these properties are KISSR\,278
(Figure~\ref{figcap6}), KISSR\,561 (Figure~\ref{figcap8}, but note
that the compact and high surface brightness object in the northern
part of the stellar disk is a foreground star), and KISSR\,572
(Figure~\ref{figcap9}).

The images shown in Figures~\ref{figcap3} through \ref{figcap10}
reveal that five of the eight neutral gas disks extend well beyond the
beam size at this sensitivity level; the \HI\ disks of KISSR\,278,
KISSR\,396, and KISSR\,572 are the same size as the beam.  There are
low surface brightness extensions of \HI\ gas in some systems:
IC\,2271 (Figure~\ref{figcap3}) and KISSR\,1021
(Figure~\ref{figcap10}) harbor \HI\ disks that are roughly twice the
size of the stellar body; KISSR\,245 (Figure~\ref{figcap5}) and
KISSR\,561 (Figure~\ref{figcap8}) show tenuous \HI\ filaments
extending toward the southwest.  However, the low signal to noise
ratio of most of these datasets (see discussion in \S~\ref{S2})
precludes a robust kinematic study of seven of the eight systems in
this work.  The detailed dynamical analysis of \adbs\ is already
presented in \citet{cannon09}, to which we refer the interested reader
for details and results.

While dynamical analyses are not possible with most of the data
presented here, the derived surface densities are robustly measured
(at this modest sensitivity level).  Stated differently, an analysis
of the (projected) surface densities in each system does not require a
detailed knowledge of the gas dynamics.  We thus limit our
interpretation to those qualities and quantities of the neutral gas
that do not depend on dynamical properties.  We focus on quantifying
the relative sizes of the neutral gas and stellar disks, as a measure
of the frequency of the ``giant disk'' phenomenon; most of \S~\ref{S4}
is devoted to addressing this science goal.

One such well-studied metric is the empirical relationship between
\HI\ column density and ongoing star formation.  While we do not have
H$\alpha$ imaging of all of these sources, each is a spectroscopically
confirmed emission line source from KISS
\citep{salzer00,salzer01,salzer02}.  We make the implicit assumption
that the majority of the ongoing star formation in these systems is
spatially coincident with the high surface brightness stellar
components shown in the optical images in Figures~\ref{figcap3}
through \ref{figcap10}.  We note explicitly that the most extreme
member of this sample, \adbs, has an H$\alpha$ morphology that is
essentially indistinguishable from that of the r-band image shown in
Figure~\ref{figcap4} (see, e.g., Figure 5 of {Cannon
  \etal\ 2009}\nocite{cannon09}).

A gas column density of 10$^{21}$ cm$^{-2}$ has been shown to trace
locations of active star formation (e.g., Skillman
1987\nocite{skillman87}; Kennicutt 1989\nocite{kennicutt89}; Kennicutt
1998\nocite{kennicutt98}; and references therein).  Interestingly, in
all of the galaxies, except IC 2271 and KISSR\,561, the peak observed
\HI\ column densities are below this 10$^{21}$ cm$^{-2}$ level, even
when the \HI\ column density and stellar surface brightness maxima are
spatially coincident.  This is consistent with the results of other
studies of dIrr systems
\citep[e.g.,][]{hunter96,meurer96,vanzee97,hunter98}.  It is important
to note that beam smearing may be removing the signatures of high
column density gas at our current resolutions of 15\arcsec\ and
20\arcsec, so we are cautious not to over-interpret this signature at
the present time.  Higher resolution and sensitivity observations of
these galaxies, while requiring extensive amounts of additional
observing time, would be very insightful in revealing the small-scale
structure of the neutral gas disks.

\subsection{Objects with Companions}
\label{S3.2}

The enigmatic properties of \adbs\ led \citet{cannon09} to argue for
an evolutionary scenario where isolation allows for a very inefficient
mode of star formation or, alternatively, very inefficient gas
consumption.  One might naively expect that systems with similar
optical properties might also be similarly isolated.  We were thus
somewhat surprised to discover that the majority of the sample members
(five of eight systems) are in relative physical proximity to other
galaxies; we hereafter refer to these systems as loose associations.
\HI\ and optical images of these loose associations are shown in
Figure~\ref{figcap11} (IC 2271), Figure~\ref{figcap12} (KISSR\,245),
Figure~\ref{figcap13} (KISSR\,396), Figure~\ref{figcap14}
(KISSR\,561), and Figure~\ref{figcap15} (KISSR\,572).  Using the 10
kpc physical length denoted in panel (b) of each of these figures, one
can discern the (projected) separation of the galaxies in each loose
association.  We draw attention to the (projected) separation of each
association member from the compact dwarf system studied here in the
caption of each figure and in Table~\ref{t5}; the minimum and
maximum separations are 26 kpc (IC\,2271 and
SDSS\,J081808.31+243006.4) and 230 kpc (KISSR\,572 and SDSS
J144754.68+291929.4).

A commonly posited mechanism to initiate a burst of star formation
(i.e., a starburst) is a tidal interaction.  Dynamical instabilities
can drive gas deep into the potential well, initiating spatially and
temporally concentrated star formation
\citep[e.g.,][]{hernquist89,mihos94}.  While a direct signature of a
tidally triggered star formation event has been difficult to produce
observationally, one might suspect that extended neutral gas in the
form of tidal tails might be considered to be an indirect signature of
interaction-driven evolution \citep[e.g., the extended tidal structure
  in the M81 group of galaxies;][]{yun94}.

To search for such signs of interactions amongst the loose
associations studied here, we examined the moment zero (integrated
\HI\ intensity) maps and full datacubes at a variety of spatial
resolutions (from 15\arcsec\ to 60\arcsec) for signatures of tidal
material between the compact dwarf systems and other association
members.  Some systems have asymmetric, low surface brightness
extensions of \HI\ gas away from the central disk (see discussion of
IC\,2271, KISSR\,1021, KISSR\,245, and KISSR\,561 above) that can be
interpreted as weak evidence for gravitational interactions playing a
role in their recent evolution.  None of the loose associations shown
in Figures~\ref{figcap11} through \ref{figcap15} show direct evidence
of strong tidal interaction in the form of tails or diffuse gas (at
the N$_{\rm HI}$ $\gsim$ 10$^{19}$ cm$^{-2}$ level) connecting one
system to another.  The most promising candidate for a strong
interaction scenario might be the loose association containing
IC\,2271; there is some low surface brightness \HI\ gas surrounding
multiple galaxies in this field.  Given the lack of very short {\it
  uv} spacings in the data presented here, the surface brightness
sensitivity to large-scale structure in these images is only modest.
We conclude that the membership of the majority of these star-forming
systems in loose associations is intriguing but weak evidence for
tidally triggered star formation events; very deep observations of
these systems with the compact {\it VLA} D configuration could be very
fruitful in further quantifying the role of interactions in initiating
starburst events.

\section{Comparing the Stellar and Gaseous Disks in the Compact Dwarfs Sample}
\label{S4}

The primary goal of the present investigation is to quantify the
relative sizes of the neutral gas and stellar disks of each of the
optically compact dwarf galaxies in our sample, thus estimating the
frequency of the ``giant disk'' phenomenon exemplified by \adbs\ and
discussed in detail in \citet{cannon09}.  To make this comparison, we
create radially averaged surface brightness profiles of each system in
the optical, infrared, and \HI\ spectral line.  We measure the
exponential scale lengths at optical and infrared wavelengths, and
compare these to both the exponential scale length of the neutral gas
disks (when possible) and to the sizes of the gaseous disks at the
10$^{19}$ cm$^{-2}$ level.  It is important to emphasize that while
this comparison does not include dynamical parameters for each system
(e.g., an inclination derived from dynamical modeling of an individual
galaxy's velocity field), the comparisons discussed below do use the
same corrections (when applicable) for galaxy inclination at each
wavelength.  Similarly, our analysis uses matched apertures between
wavelengths to assure self-similar measurements.

We define a galaxy as being ``resolved'' in a certain image if the
derived scale length or size exceeds the FWHM or beam size.
Therefore, we classify systems as ``resolved'' in the {\it SDSS}
r-band, the {\it Spitzer} 3.6 $\mu$m band, or the {\it Spitzer} 4.5
$\mu$m band, if the derived scale lengths exceed 1.3\arcsec,
1.66\arcsec, or 1.72\arcsec, respectively.  Similarly, we classify
systems as ``resolved'' in \HI\ if the size of the \HI\ disk at the
10$^{19}$ cm$^{-2}$ column density level exceeds the 15\arcsec\ beam
size.  We discuss the details of each of these metrics in the
subsections that follow.

\subsection{Optical and Infrared Surface Brightness Profiles}
\label{S4.1}

Optical and infrared surface brightness profiles for each of the
galaxies were produced by fitting elliptical isophotes to the {\it
  SDSS} r-band, {\it Spitzer} 3.6 $\mu$m, and {\it Spitzer} 4.5 $\mu$m
images using the IRAF\footnote{The Image Reduction and Analysis
  Facility (IRAF) is distributed by the National Optical Astronomy
  Observatory, which is operated by AURA, Inc., under cooperative
  agreement with the National Science Foundation.} task ELLIPSE.  As
noted above, some of these systems have easily identifiable foreground
stars that can contaminate the surface brightness profiles [e.g.,
  \adbs\ (Figure~\ref{figcap4}) and KISSR\,561
  (Figure~\ref{figcap8})].  Such foreground objects were masked by
hand using the IRAF task FIXPIX.  

For each system, the position angle and inclination were determined
from the {\it SDSS} r-band image.  When a system was extended in one
dimension, the position angle was defined as that angle east of north
that aligns with the larger projected optical axis.  The inclination
was defined by \begin{math}i =
  cos^{-1}\left(\frac{b}{a}\right) \end{math} where {\it a} and {\it
  b} are the major and minor axes, respectively.  The ellipse
parameters used to derive the isophotal photometry are given in
Table~\ref{t6}.  Surface photometry was measured in elliptical
isophotes incremented in one-pixel steps along the major axis
(0.4\arcsec\ per ellipse for the {\it SDSS} images and 0.6\arcsec\ per
ellipse for the {\it Spitzer} images).  The r-band ellipses are shown
in red in Figures~\ref{figcap3} though \ref{figcap10}, while the
infrared ellipses are shown in green in Figures~\ref{figcap5} though
\ref{figcap10}.

Figures~\ref{figcap16} and \ref{figcap17} show the resulting surface
brightness plots for all eight of the sample galaxies.  As expected
based on the selection criteria used to create the sample (two of
which were an optical radius less than 3\arcsec\ and an upper limit to
the recessional velocity or distance; see discussion in \S~\ref{S2}
above), the profiles are all centrally peaked and show little
structure throughout the optical disks of the galaxies.  While there
is some variation in the steepness of the profile in the inner regions
of the systems (i.e., within the inner few arcseconds), it is
immediately obvious that each system harbors an exceptionally compact
stellar component.

To quantify the sizes of these compact stellar populations, we next
derived the scale lengths of each system in the {\it SDSS} r-band and
in the {\it Spitzer} 3.6 $\mu$m and 4.5 $\mu$m bands.  In deriving
these profiles, we follow the methodology in \citet{deblok95}.  Using 
logarithmic
units of mag arcsec$^{-2}$, a very good approximation for the
exponential disk component of the optical radial profile can be made
using:
\begin{equation}
\mu(r)=\mu_{\circ} + 1.086\left(\frac{r}{h}\right)
\label{eq1}
\end{equation}
where $\mu _{\circ}$ is the central surface brightness in units of mag
arcsec$^{-2}$, {\it r} is the radius, $\mu(r)$ is the surface
brightness in units of mag arcsec$^{-2}$ as a function of radius, and
{\it h} is the scale length.  Table~\ref{t7} provides the calculated
scale lengths (in both angular and in physical units) at each
wavelength, derived from the profiles shown in Figures~\ref{figcap16}
and \ref{figcap17}, with upper limit errors of 25\% at each ellipse
location.  Note that we use the nomenclature of \hr, \hthree, and
\hfour\ to denote the scale lengths (in angular units and in physical
units of kpc) of each galaxy in the {\it SDSS} r-band, the {\it
  Spitzer} 3.6 $\mu$m band, and the {\it Spitzer} 4.5 $\mu$m band,
respectively.

The scale lengths presented in Table~\ref{t7} show that all of the
galaxies are resolved in at least one optical or infrared image, and
that most are resolved in both.  Recall that the FWHM of the {\it
  SDSS} r-band, {\it Spitzer} 3.6 $\mu$m, and {\it Spitzer} 4.5 $\mu$m
images are 1.3\arcsec, 1.66\arcsec, and 1.72\arcsec, respectively.
The \hr\ values in Table~\ref{t7} show that all systems except for
KISSR\,245 are resolved in the {\it SDSS} images (that is,
\hr\ exceeds the 1.3\arcsec\ FWHM of the {\it SDSS} r-band images).
The \hr\ value for KISSR\,245 is formally smaller than the FWHM,
although it is consistent within the errorbar.  Similarly, {\it
  Spitzer} resolves each galaxy; note, however, that the scale lengths
of KISSR\,245 at 3.6 $\mu$m and of KISSR\,1021 at 4.5 $\mu$m are
consistent with the FWHM, given the errorbars.

The optical and infrared scale lengths in Table~\ref{t7} quantify the
sizes of the compact stellar components of the sample galaxies.  At
the adopted distances, the stellar component of each system has an
r-band scale length much less than 1 kpc.  All systems have
\hr\ $\leq$0.5 kpc, except for KISSR\,561.  Moving to the infrared,
the scale lengths increase.  This is exactly the behavior expected of
an older and non-star forming stellar population.  While each of the
infrared scale lengths exceeds the r-band value, it is remarkable to
note that each system is still exceptionally compact even when
considering the older stellar populations.  For example, at 4.5
$\mu$m, some systems have scale lengths slightly larger than 1 kpc
(KISSR\,278, KISSR\,561, KISSR\,572), but others show very little
increase in scale length with increasing wavelength (consider
KISSR\,1021, whose scale length does not change between the optical
and the infrared, within the measurement errors).

\subsection{\HI\ Surface Brightness Profiles}
\label{S4.2}

Having quantitatively established the compact nature of the stellar
population of each of the sample galaxies, we now seek to quantify the
sizes of the \HI\ neutral gas disks via two methods.  First, we derive
radially averaged surface brightness profiles for each system; when
possible, we use these profiles to derive an exponential scale length.
Second, we quantify the size of the disks at the 10$^{19}$ cm$^{-2}$
levels.  We discuss each method in detail below.

We begin by creating radial \HI\ surface brightness profiles for each
of the galaxies using the 15\arcsec\ beam images and the
GIPSY\footnote{The Groningen Image Processing System.} task ELLINT.
As an inspection of Figures~\ref{figcap3} through \ref{figcap10}
shows, the neutral gas disks studied here are larger than the beam
size for five of the eight galaxies (excluding KISSR\,278, KISSR\,396,
and KISSR\,572).  A detailed dynamical analysis is only possible for
\adbs; this precludes the derivation of kinematic position angles and
inclinations for most galaxies.  In order to assure an accurate
comparison of stellar and gaseous properties, we thus use the same
position angles and inclinations as derived using the {\it SDSS}
r-band images (see Table~\ref{t6} and the detailed discussion in
\S~\ref{S4.1} above).  These ellipses are shown in blue in
Figures~\ref{figcap3} through \ref{figcap10}.  A semi-major axis and
step size of 5\arcsec\ was used to create each of the \HI\ profiles;
technically this means that the profiles we discuss below are
oversampled (compared to the 15\arcsec\ beam size) by a factor of 3.
However, this oversampling does allow us to extract meaningful
radially averaged surface brightness profiles.

Figures 18 and 19 show these surface brightness profiles for each of
the eight sample galaxies.  As was evident from the surface density
images presented in \S~\ref{S3.1}, most of the sample galaxies have
\HI\ surface brightness profiles that are centrally peaked at the
resolution of these observations and that fall off with radius in a
more or less smooth way.  Note that low surface brightness features in
some of the galaxies' moment zero (integrated \HI\ intensity) maps do
cause departures from smooth profiles in the outer disks; for example,
the low surface brightness feature extending to the southwest of
KISSR\,245 in Figure~\ref{figcap5} is easily identified as the slight
increase in \HI\ surface density between $\sim$25\arcsec\ and
$\sim$50\arcsec\ in Figure~\ref{figcap18}.

The obvious exception to this trend is \adbs, whose radially averaged
\HI\ surface density profile (created with the optically derived
inclination and position angle discussed above, and shown in red in
Figure~\ref{figcap18}) rises with increasing galactocentric radius,
peaks, and then falls off in the outer disk.  The peak surface
densities are located in the broken ring structure $\sim$7.5 kpc from
the stellar component.  A cursory comparison of the profiles in
Figures~\ref{figcap18} and \ref{figcap19} already reveals the striking
differences between \adbs\ and the other compact dwarf galaxies in the
present sample; we discuss these differences in more detail below, and
refer the interested reader to \citet{cannon09} for a more complete
discussion of the remarkable characteristics of the \adbs\ system.

The radial profiles shown in Figures~\ref{figcap18} and \ref{figcap19}
(specifically, the red profile for ADBS 113845+2008) were created
using exactly the same inclinations and position angles as used to
derive the optical and infrared surface brightness profiles in
Figures~\ref{figcap16} and \ref{figcap17}.  Using these neutral gas
radial profiles, exponential fits were made using a version of
Equation~\ref{eq1} converted to linear units:
\begin{equation}
\Sigma(r)=\Sigma_{\circ} exp\left(-\frac{r}{h}\right)
\label{eq2}
\end{equation}
where $\Sigma_{\circ}$ is the central surface brightness of the disk
in \msun\ pc$^{-2}$, {\it r} is the radius, $\Sigma(r)$ is the surface
brightness of the disk in \msun\ pc$^{-2}$ as a function of radius,
and {\it h} is the scale length \citep{deblok95}.  We hereafter refer
to the \HI\ scale length derived using this equation as \hhi.  We
tabulate the derived \hhi\ values in Table~\ref{t7}. For all of the
systems except \adbs, the calculated values of \hhi\ are smaller than
the beam size.  Formally, this metric to measure the size of the
\HI\ disks only provides meaningful results when it can be fit over
multiple independent beams.  As noted above, our radially averaged
profiles required over-sampling by a factor of three.  While the sizes
of the \HI\ disks based on the \hhi\ metric are interesting, and we
discuss these briefly below, we seek an alternative measure of the
size of the neutral gas disks of these galaxies.

It is obvious from the neutral gas morphology (Figure~\ref{figcap3})
and surface brightness profile (Figure~\ref{figcap18}) of \adbs\ that
an exponential scale length is undefined for a disk whose morphology
is not centrally concentrated.  While we considered fitting the outer
disk of \adbs\ (i.e., that region at radii larger than that of the
broken \HI\ ring) with an exponential function, the striking disparity
between the neutral gas and stellar disks of this system warrants
caution against such an approach.  Due to the remarkable qualities of
this galaxy's neutral gas disk, we thus do not calculate an
exponential \HI\ scale length.  The properties of \adbs\ naturally
require a second metric for the size of the neutral gas disks of these
systems.

To address these issues, we thus use the radially averaged
\HI\ surface density profiles in Figures~\ref{figcap18} and
\ref{figcap19} to determine the (radially averaged) size of the
\HI\ disk at the 10$^{19}$ cm$^{-2}$ column density level.  This
column density corresponds to a mass surface density of 0.08
\msun\ pc$^{-2}$; recall from the discussion in \S~\ref{S2.2} that
this emission is above the 2\,$\sigma$ level in at least 3 consecutive
channels.  We extract the radii of the sample galaxies at this level
(hereafter referred to as \rhi) from Figures~\ref{figcap18} and
\ref{figcap19} and tabulate the values in Table~\ref{t7}; note that
\citet{cannon09} quantified the physical size of the \adbs\ disk in
the same manner.  As expected, \rhi\ $>$ \hhi\ in all systems where
both quantities are defined.  It is important to stress that for
KISSR\,245, KISSR\,278, KISSR\,396, and KISSR\,572, \rhi\ is equal
(within errors) to the beam size; this means that the \rhi\ metric for
these galaxies is formally an upper limit (but note that KISSR\,245
does have \HI\ gas extending to the southwest, well beyond a single
beam size).  These four systems are ``unresolved'' by the present
observations (that is, \rhi\ $\lsim$ beam size, thus the \rhi\ and
related values being listed as upper limits in Tables~\ref{t7} and
\ref{t8}); they may harbor smaller neutral gas components than
calculated here.  We discuss the ratios of both \hhi/\hr\ and
\rhi/\hr\ for all eight sample galaxies in \S~\ref{S4.3}.

We again note that no kinematic constraints are used to derive the
scale lengths or disk sizes discussed here; the profiles shown in
Figures~\ref{figcap16} through \ref{figcap19} use only
optically-derived position angles and inclinations.  Since seven of
the eight galaxies are not sufficiently resolved to undertake a robust
dynamical analysis, we can only test these dependences for the one
system that does have a full kinematic model, \adbs.  To test how
using the position angles and inclinations derived from the optical
images affect our calculations (as opposed to using those derived from
the \HI\ dynamics), we show three radial mass surface density profiles
for \adbs\ in Figure~\ref{figcap18}.  The red profile is derived using
the optical inclination (0$^{\circ}$) and position angle (0$^{\circ}$)
and is discussed above.  The blue profile uses the dynamically-derived
position angle (155$^{\circ}$) and inclination (30$^{\circ}$) from
\citet{cannon09}.  The green profile uses the same position angle
(155$^{\circ}$) as in \citet{cannon09}, but assumes a larger value for
the galaxy's inclination angle (60$^{\circ}$).  Since \hhi\ is
undefined for \adbs, it is difficult to use this test to assess the
dependence of \HI\ scale length on inclination or position angle.  The
\rhi\ values, on the other hand, do change slightly for these
different fits (\rhi\ $=$ 17.3 $\pm$ 3.5 kpc, 18.1 $\pm$ 3.6 kpc, and
22.3 $\pm$ 4.5 kpc, for the red, blue, and green profiles in
Figure~\ref{figcap18}, respectively), but the resulting \HI\ disk
radii are equal within the (substantial) errorbars.  While differences
from one system to another are expected, this simple test gives a
baseline for how to interpret the measurements made using the position
angles and inclinations derived from the {\it SDSS} r-band images
within the rest of the sample.  We consider the \rhi\ values for
IC\,2271, \adbs, KISSR\,561, and KISSR\,1021 to be accurate
representations of their (average) physical size at the 10$^{19}$
cm$^{-2}$ level.  The \rhi\ values in Table~\ref{t7} for the other
four systems should be considered upper limits on the physical size of
each neutral gas disk.

When available, we also tested how the disk size measurements change
when using C configuration versus D configuration data.  This
comparison can only be made for \adbs\ and IC\,2271.  The scale length
measurement of IC\,2271 with the lower resolution D configuration data
is \hhi=2.90\,$\pm$\,0.73 kpc (20.0\arcsec\,$\pm$\,5.0\arcsec); this
can be compared with the \hhi=1.72\,$\pm$\,0.43 kpc
(11.9\arcsec\,$\pm$\,3.0\arcsec) value derived from the C
configuration data only (see Table~\ref{t7}).  The radii of the galaxy
disks, when using the C versus D configuration datasets, change from
\rhi=8.74\,$\pm$\,2.19 kpc to 11.6\,$\pm$\,2.9 kpc (IC\,2271) and from
17.3\,$\pm$\,4.3 kpc to 21.6\,$\pm$\,5.4 kpc (\adbs).  While these
increases are slightly larger than the sizes of the errorbars, we
conclude that there is only a modest dependency of the size estimates
of these galaxy disks on the use of only the C configuration data.
Given that our analysis uses the same imaging strategy for all
systems, the surface brightness sensitivity of each dataset is very
comparable.

\subsection{On the Frequency of the ``Giant Disk'' Phenomenon in Compact 
Dwarf Galaxies}
\label{S4.3}

We can now compare, in a quantitative sense, the relative sizes of the
stellar and gaseous disks of the eight galaxies in this sample.
Table~\ref{t8} contains the (dimensionless) ratios of \hhi\ and
\rhi\ versus \hr, \hthree, and \hfour, for all galaxies with available
data (recall that \adbs\ has an undefined \hhi, and that no {\it
  Spitzer} imaging is available for \adbs\ or for IC\,2271).  From the
contents of this table, we note the following trends and behaviors.

As discussed in \S~\ref{S4.2} above, four of the eight systems have
\HI\ disks that exceed the sizes of the stellar bodies based on the
\rhi\ metric (and are thus considered ``resolved'' in \HI).  Hence,
the dimensionless ratios in Table~\ref{t8} for KISSR\,245, KISSR\,278,
KISSR\,396, and KISSR\,572 should be considered upper limits; the
actual size ratios of neutral to stellar disks may in fact be smaller
for these systems.  For the entire sample, the ratios of \hhi\ to
r-band or infrared scale lengths (spanning a range of $\sim$2 to
$\sim$7) are systematically smaller than the ratios of \rhi\ to the
same r-band or infrared values (which span a range of $\sim$4 to
$\sim$58).  Similarly, for each individual galaxy, the relative size
ratios measured using either \rhi\ or \hhi\ decrease with increasing
wavelength used to measure the stellar disk size.  These general
trends hold across the entire sample, regardless of which tracer is
used to measure the size of either the stellar or the gaseous disk.

The \rhi/\hr\ ratio for \adbs\ (58\,$\pm$\,21) is more than a factor
of two larger than the same ratio for any of the other systems; if the
ratios in Table~\ref{t7} for the other sample members are smaller,
then this reinforces the enormity of the \HI\ disk of \adbs.  Using
the other seven sample galaxies as benchmarks, it is physically
intuitive that the infrared scale length of \adbs\ would be larger
than the measured \hr\ value.  However, the enormous \rhi/\hr\ ratio
of this system, and the relative size of the broken \HI\ ring
($\sim$7.5 kpc in radius) versus the optical scale length
(0.30\,$\pm$\,0.08 kpc), are challenging parameters to explain unless
the optical body is only a high surface brightness knot in an
otherwise extremely low surface brightness underlying stellar disk.
We again refer the interested reader to \citet{cannon09} for more
detailed discussion; for the purposes of this work, we interpret the
remarkable properties of \adbs\ as the signature of a ``giant''
\HI\ disk.

The present sample of compact dwarf galaxies was created using
consistent selection criteria; the observations and analysis have
produced a uniform set of results.  While the sample is small, it
appears that the frequency of ``giant disk'' dwarf galaxies such as
\adbs\ is low.  A few other examples of such systems have been
identified (e.g., DDO\,154, {Carignan \& Beaulieu
  1989}\nocite{carignan89}, {Carignan \& Purton
  1998}\nocite{carignan98}; NGC\,2915, {Meurer
  \etal\ 1996}\nocite{meurer96}; UGC\,5288, {van~Zee
  2004}\nocite{vanzee04}; NGC\,3741, {Bremnes
  \etal\ 2000}\nocite{bremnes00}, {Begum
  \etal\ 2005}\nocite{begum05}).  This small collection of galaxies
suggests that systems with qualities similar to \adbs\ are
comparatively rare in the local volume.  A larger sample of
\HI\ observations of local compact dwarf systems could be fruitful in
improving the statistics of the trends identified in Table~\ref{t8}.

\section{Conclusions}
\label{S5}

We have presented {\it VLA} C configuration imaging of eight optically
compact dwarf galaxies.  These sources were selected to share the
following common characteristics: 1) compact stellar populations; 2)
faint blue absolute magnitudes (i.e., low masses); 3) emission lines
indicative of recent star formation; and 4) proximity for spatially
resolved work.  The sample includes the ``giant disk'' dwarf galaxy
\adbs\ studied in \citet{cannon09}.  The {\it VLA} data are
supplemented by archival optical ({\it SDSS} r-band) and infrared
({\it Spitzer} 3.6 $\mu$m and 4.5 $\mu$m) imaging.  We use this data
suite to quantitatively address the relative sizes of the stellar and
gaseous components of these galaxies.

A comparison of the morphology of the stellar and gaseous disks of
each galaxy in the sample reveals that all eight systems have
extremely compact (though in almost all cases resolved) stellar
components; this is of course expected given the selection criteria.
Seven of the eight systems in the sample harbor centrally concentrated
neutral gas distributions.  \adbs\ differs from the other systems in
that its neutral gas disk is structured in a broken ring that is well
outside of the stellar component. In most of the other cases, the
stellar and gaseous disks have surface brightness maxima that are in
close physical proximity to one another.

Five of the eight systems are located in loose associations, with
neighboring systems with projected separations of less than a few
hundred kpc and with recessional velocities within a few hundred
km\,s$^{-1}$. Some of the loose association galaxies have low surface
brightness neutral gas in the outer disks that might be construed as
evidence of tidal interactions.  However, given the lack of very short
{\it uv} spacings in our data, we conservatively conclude that the
membership of these star-forming dwarfs in such loose associations is
only weak evidence for a tidal burst trigger.

While each of the eight sample galaxies is detected in \HI\ at
meaningful significance, these {\it VLA} data do not allow a robust
dynamical analysis in any of the systems except for \adbs.  Thus, we
concentrate on the morphological properties of the galaxies.  We use
the optical, infrared, and \HI\ data to derive radially averaged
surface brightness profiles that can be used to measure the sizes of
both the stellar and the gaseous components.  We quantify the sizes of
the \HI\ disks via both an exponential scale length (available for all
systems except for \adbs) and a (radially averaged) limit of the
gaseous disk at the 10$^{19}$ cm$^{-2}$ column density (0.08
\msun\ pc$^{-2}$) level. 

\begin{figure}
\epsscale{1.0}
\plotone{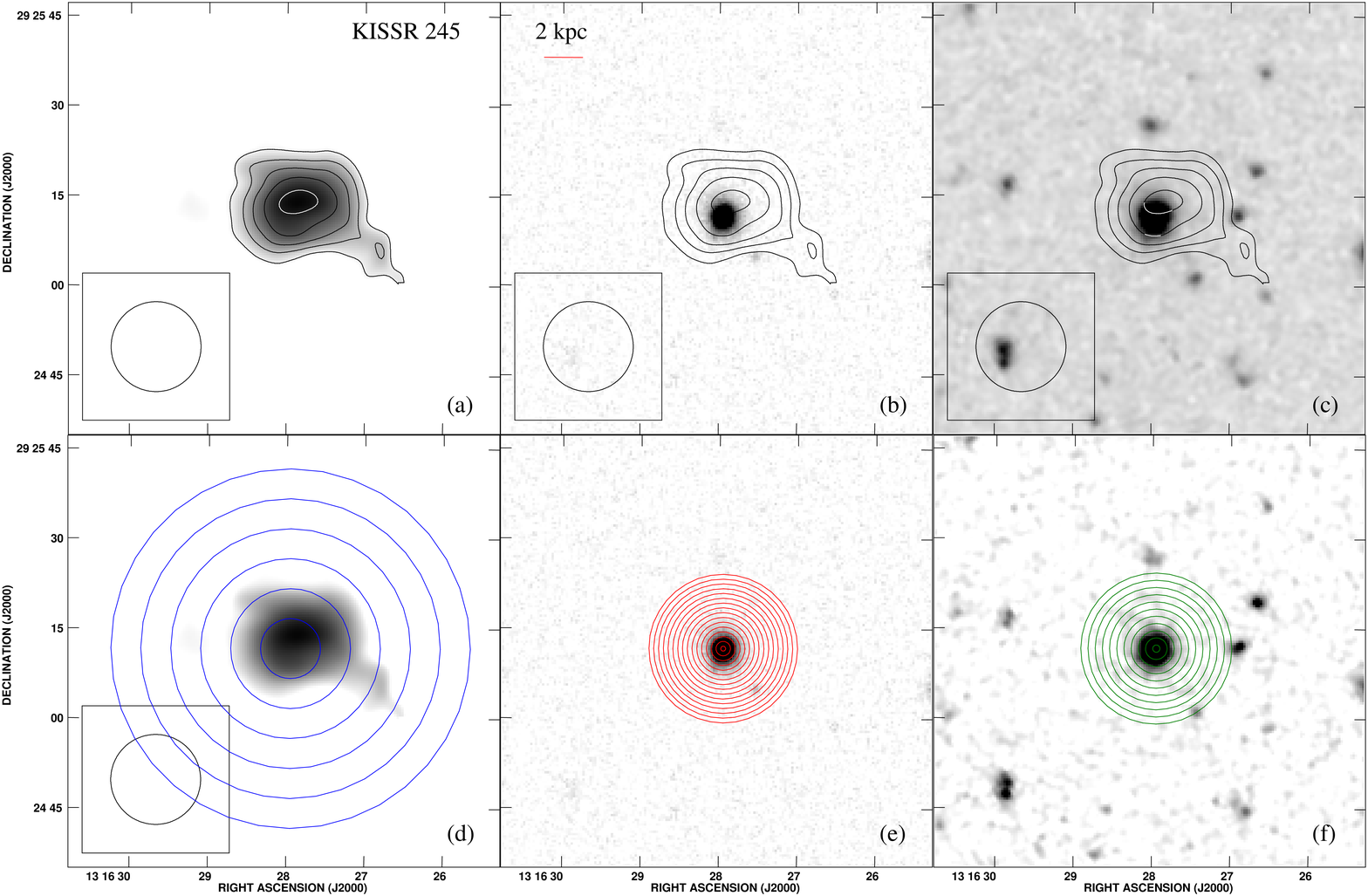}
\epsscale{1.0}
\caption{{\it VLA}, {\it SDSS}, and {\it Spitzer} images of
  KISSR\,245.  Panels (a) and (d) show the \HI\ column density at
  15\arcsec\ resolution; panels (b) and (e) show the {\it SDSS} r-band
  image; panels (c) and (f) show the Spitzer 3.6 $\mu$m (c) and 4.5
  $\mu$m (f) images. The contours in panels (a) through (c) are at the
  (0.8, 1, 1.2, 1.4, 1.6)\,$\times$\,10$^{20}$ cm$^{-2}$ levels.
  Panel (d) is overlaid with the isophotal ellipses used to create the
  \HI\ radial profile (see \S~\ref{S4.2}), while panels (e) and (f)
  are overlaid with the isophotal ellipses used to create the optical
  and infrared radial profiles (see \S~\ref{S4.1}). The ellipses in
  panels (e) and (f) are shown as every other to be visible at this
  size.  The red bar in panel (b) denotes a physical length of 2 kpc
  at the adopted distance.}
\label{figcap5}
\end{figure}

\begin{figure}
\epsscale{1.0}
\plotone{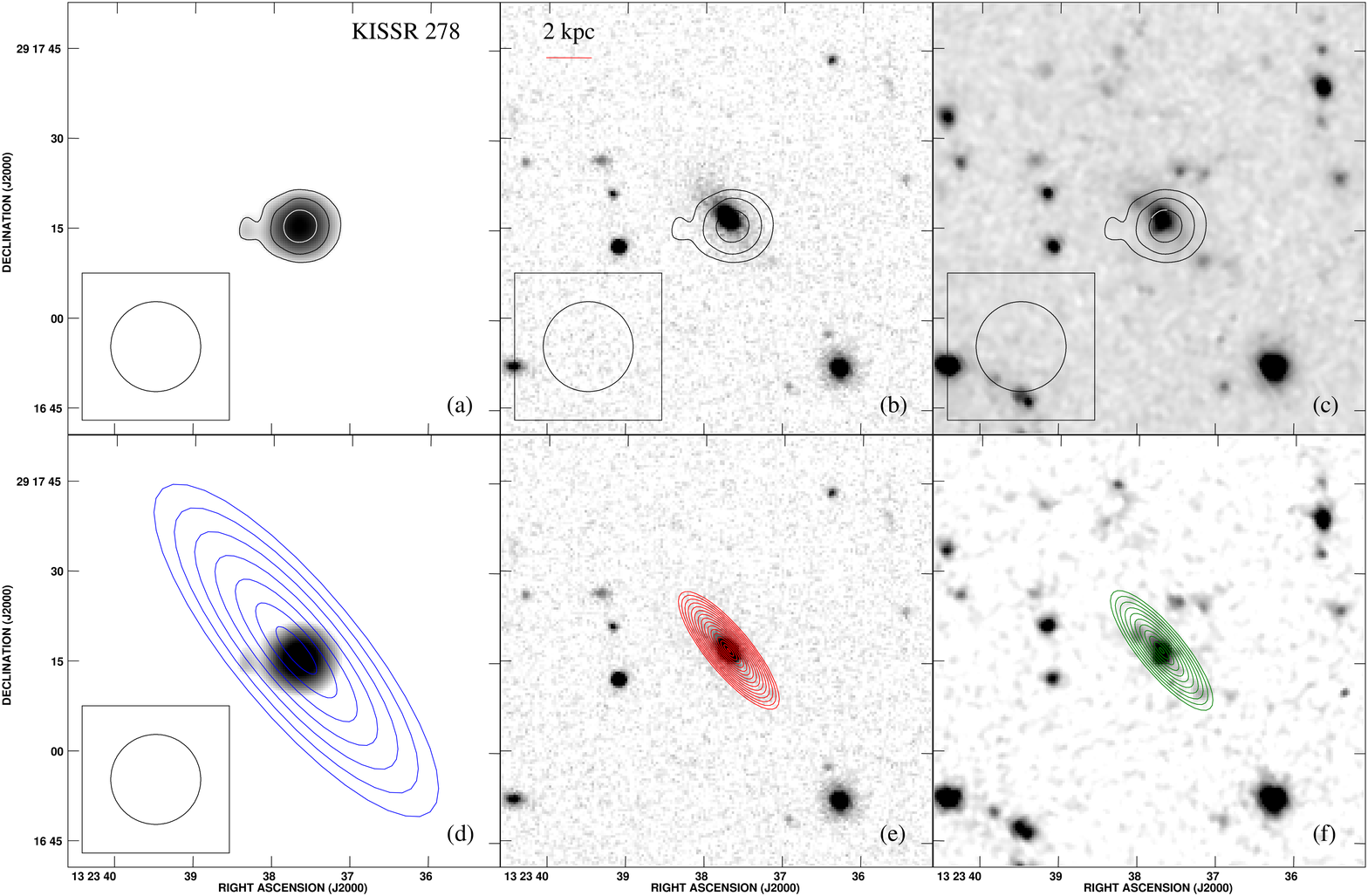}
\epsscale{1.0}
\caption{{\it VLA}, {\it SDSS} and {\it Spitzer} images of KISSR\,278.
  Panels (a) and (d) show the \HI\ column density at
  15\arcsec\ resolution; panels (b) and (e) show the {\it SDSS} r-band
  image; panels (c) and (f) show the {\it Spitzer} 3.6 $\mu$m (c) and
  4.5 $\mu$m (f) images. The contours in panels (a) through (c) are at
  the (0.8, 1, 1.2)\,$\times$\,10$^{20}$ cm$^{-2}$ levels.  Panel (d)
  is overlaid with the isophotal ellipses used to create the
  \HI\ radial profile (see \S~\ref{S4.2}), while panels (e) and (f)
  are overlaid with the isophotal ellipses used to create the optical
  and infrared radial profiles (see \S~\ref{S4.1}).  The ellipses in
  panels (e) and (f) are shown as every other to be visible at this
  size.  The red bar in panel (b) denotes a physical length of 2 kpc
  at the adopted distance.  }
\label{figcap6}
\end{figure}

\begin{figure}
\epsscale{1.0}
\plotone{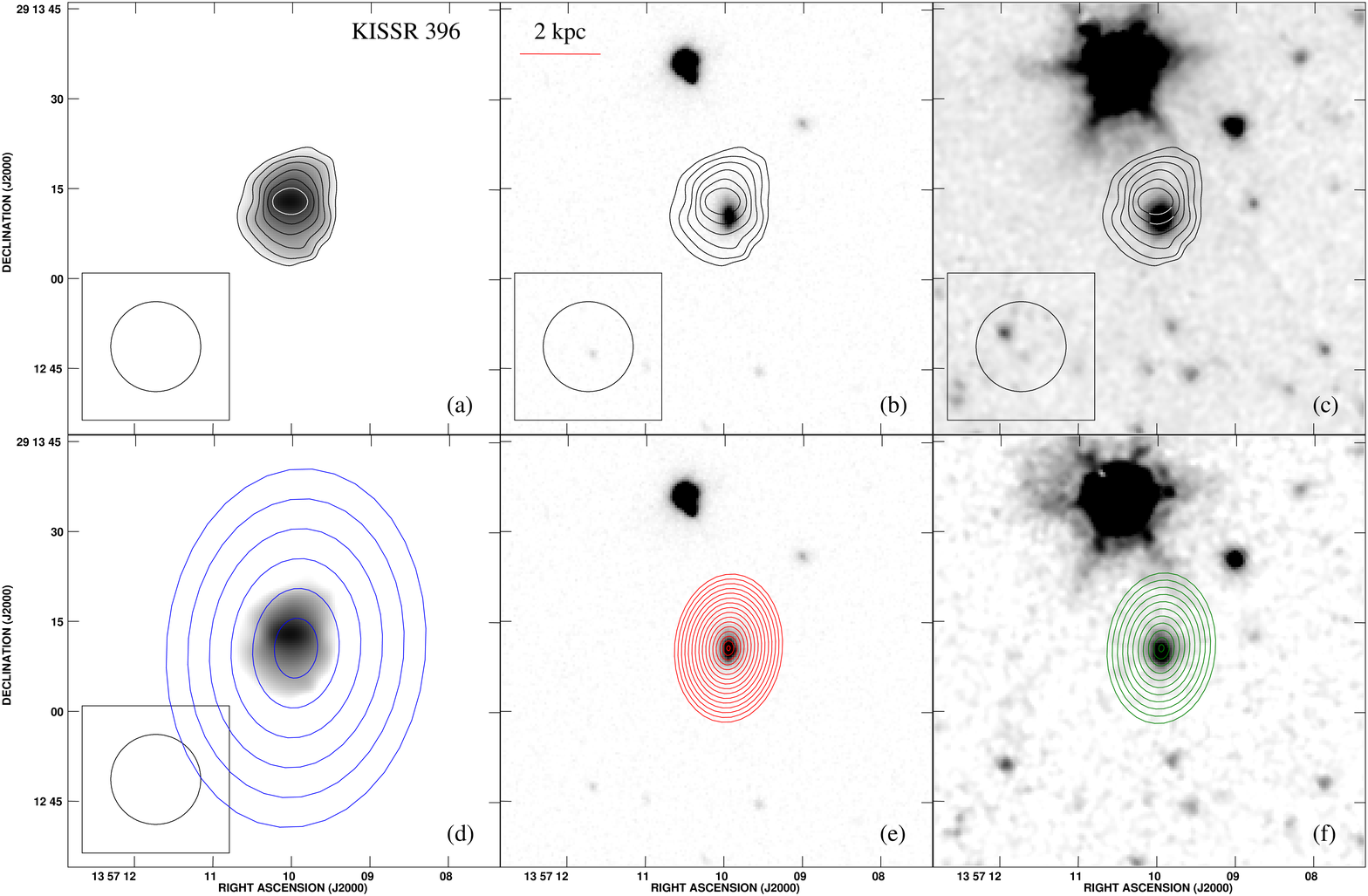}
\epsscale{1.0}
\caption{{\it VLA}, {\it SDSS} and {\it Spitzer} images of KISSR\,396.
  Panels (a) and (d) show the \HI\ column density at
  15\arcsec\ resolution; panels (b) and (e) show the {\it SDSS} r-band
  image; panels (c) and (f) show the {\it Spitzer} 3.6 $\mu$m (c) and
  4.5 $\mu$m (f) images. The contours in panels (a) through (c) are at
  the (0.8, 1, 1.2, 1.4, 1.6, 1.8)\,$\times$\,10$^{20}$ cm$^{-2}$
  levels.  Panel (d) is overlaid with the isophotal ellipses used to
  create the \HI\ radial profile (see \S~\ref{S4.2}), while panels (e)
  and (f) are overlaid with the isophotal ellipses used to create the
  optical and infrared radial profiles (see \S~\ref{S4.1}).  The
  ellipses in panels (e) and (f) are shown as every other to be
  visible at this size.  The red bar in panel (b) denotes a physical
  length of 2 kpc at the adopted distance.  }
\label{figcap7}
\end{figure}

\begin{figure}
\epsscale{1.0}
\plotone{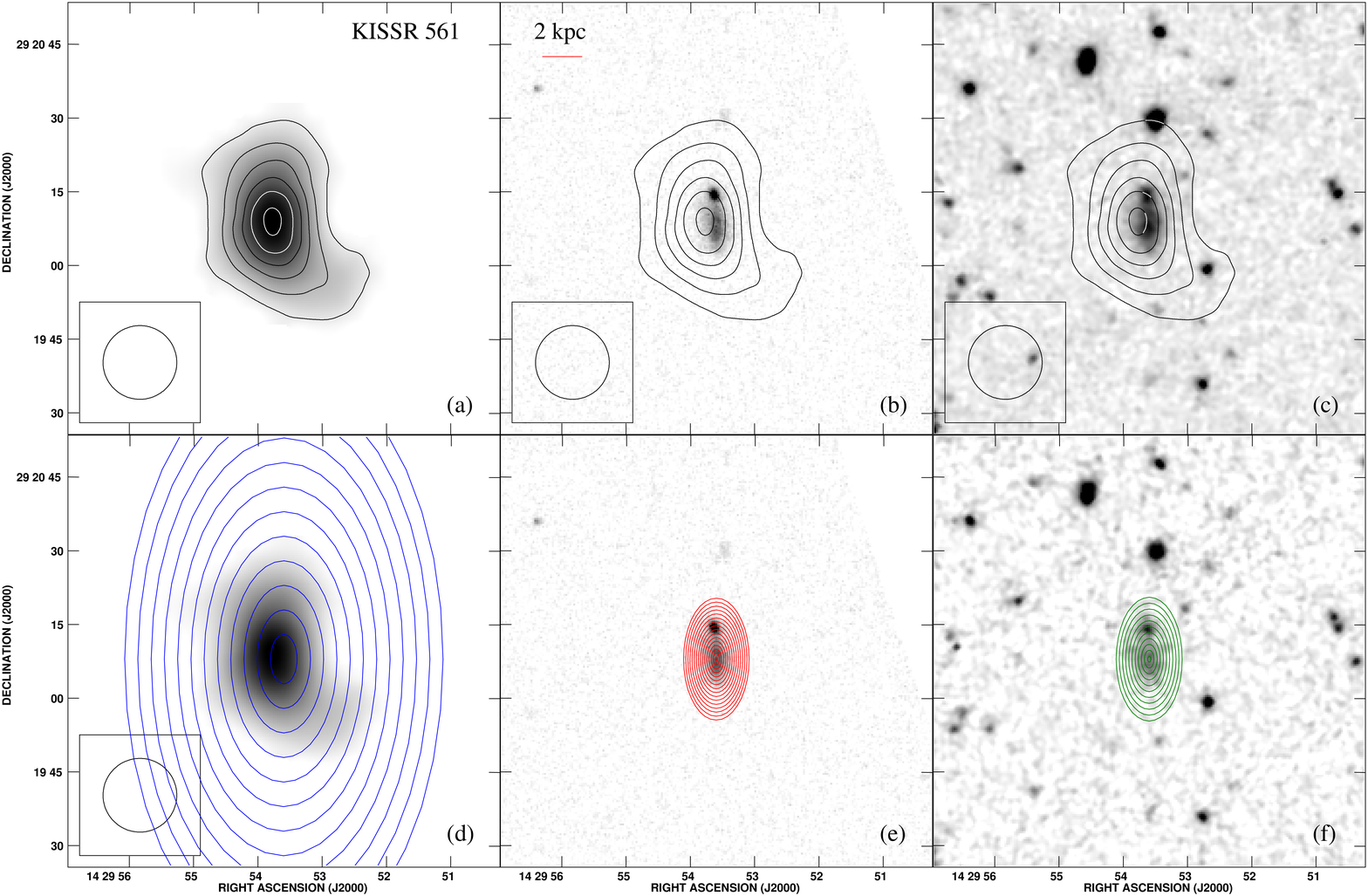}
\epsscale{1.0}
\caption{{\it VLA}, {\it SDSS} and {\it Spitzer} images of KISSR\,561.
  Panels (a) and (d) show the \HI\ column density at
  15\arcsec\ resolution; panels (b) and (e) show the {\it SDSS} r-band
  image; panels (c) and (f) show the {\it Spitzer} 3.6 $\mu$m (c) and
  4.5 $\mu$m (f) images. The contours in panels (a) through (c) are at
  the (2, 4, 6, 8, 10, 12)\,$\times$\,10$^{20}$ cm$^{-2}$ levels.
  Panel (d) is overlaid with the isophotal ellipses used to create the
  \HI\ radial profile (see \S~\ref{S4.2}), while panels (e) and (f)
  are overlaid with the isophotal ellipses used to create the optical
  and infrared radial profiles (see \S~\ref{S4.1}).  The ellipses in
  panels (e) and (f) are shown as every other to be visible at this
  size.  The red bar in panel (b) denotes a physical length of 2 kpc
  at the adopted distance.  }
\label{figcap8}
\end{figure}

\begin{figure}
\epsscale{1.0}
\plotone{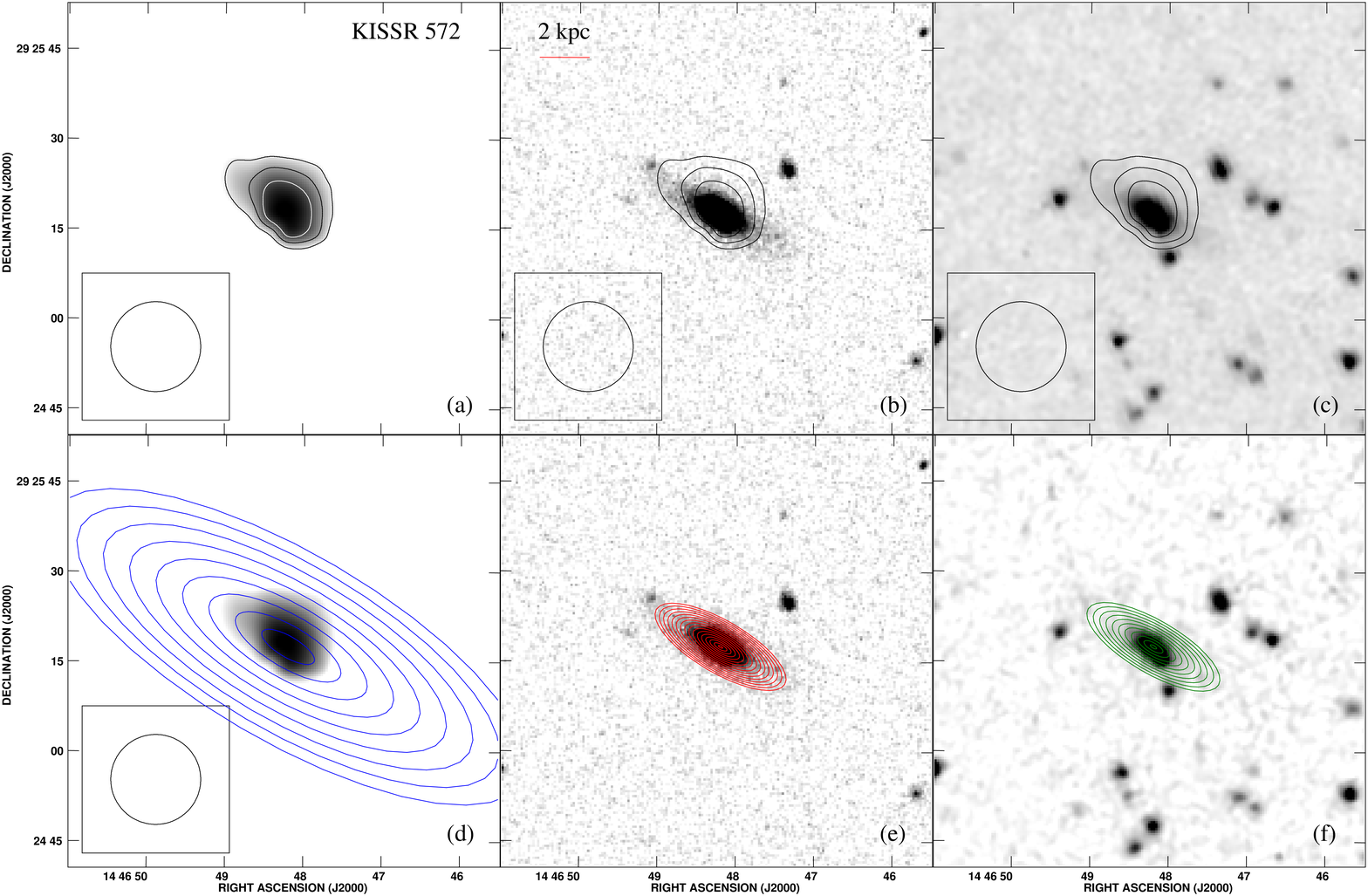}
\epsscale{1.0}
\caption{{\it VLA}, {\it SDSS} and {\it Spitzer} images of
  KISSR\,572. Panels (a) and (d) show the \HI\ column density at
  15\arcsec\ resolution; panels (b) and (e) show the {\it SDSS} r-band
  image; panels (c) and (f) show the {\it Spitzer} 3.6 $\mu$m (c) and
  4.5 $\mu$m (f) images. The contours in panels (a) through (c) are at
  the (0.8, 1, 1.2)\,$\times$\,10$^{20}$ cm$^{-2}$ levels.  Panel (d)
  is overlaid with the isophotal ellipses used to create the
  \HI\ radial profile (see \S~\ref{S4.2}), while panels (e) and (f)
  are overlaid with the isophotal ellipses used to create the optical
  and infrared radial profiles (see \S~\ref{S4.1}).  The ellipses in
  panels (e) and (f) are shown as every other to be visible at this
  size. The red bar in panel (b) denotes a physical length of 2 kpc at
  the adopted distance.  }
\label{figcap9}
\end{figure}

\begin{figure}
\epsscale{1.0}
\plotone{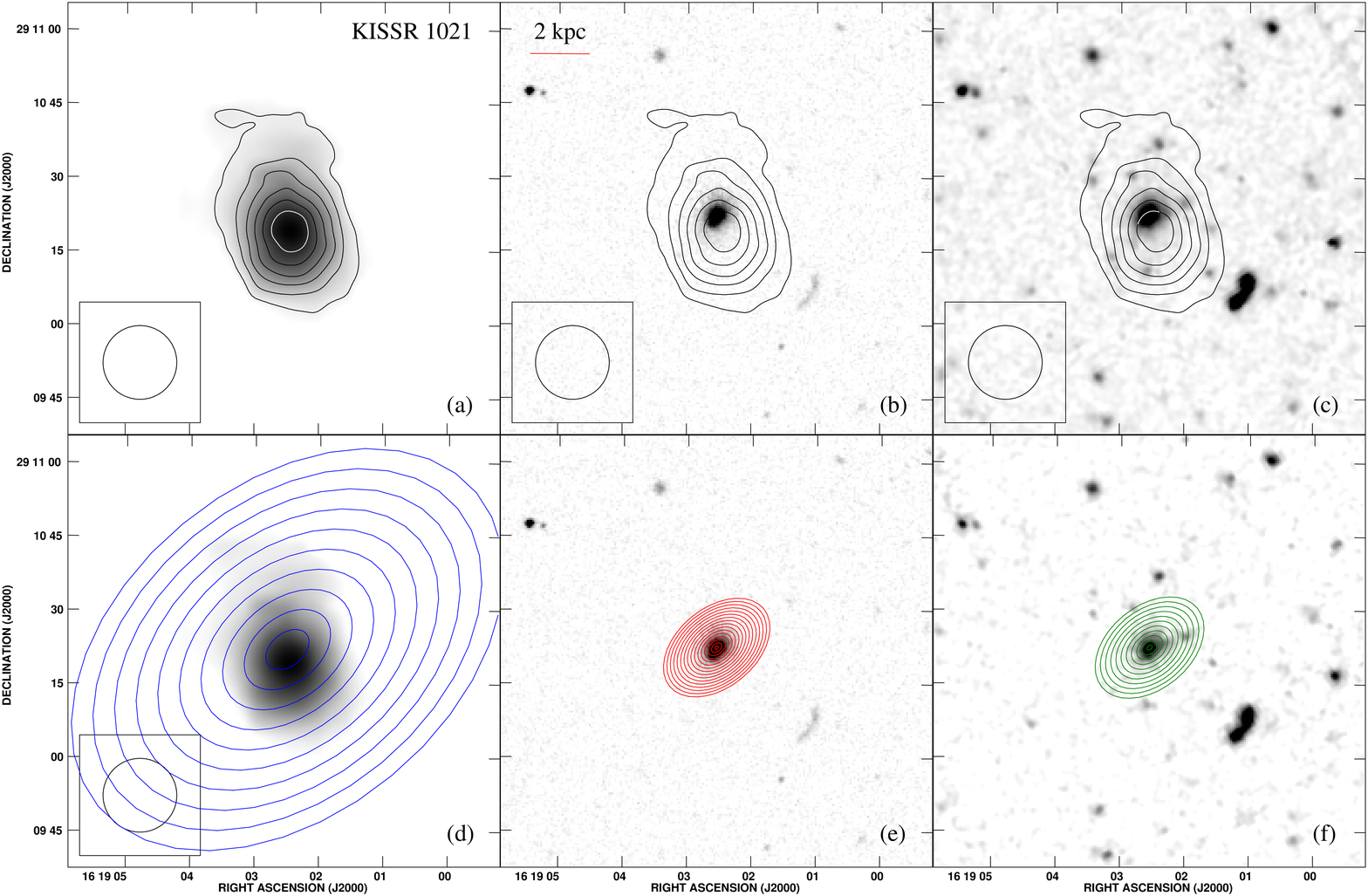}
\epsscale{1.0}
\caption{{\it VLA}, {\it SDSS} and {\it Spitzer} images of
  KISSR\,1021.  Panels (a) and (d) show the \HI\ column density at
  15\arcsec\ resolution; panels (b) and (e) show the {\it SDSS} r-band
  image; panels (c) and (f) show the {\it Spitzer} 3.6 $\mu$m (c) and
  4.5 $\mu$m (f) images. The contours in panels (a) through (c) are at
  the (1, 2, 3, 4, 5, 6)\,$\times$\,10$^{20}$ cm$^{-2}$ levels.  Panel
  (d) is overlaid with the isophotal ellipses used to create the
  \HI\ radial profile (see \S~\ref{S4.2}), while panels (e) and (f)
  are overlaid with the isophotal ellipses used to create the optical
  and infrared radial profiles (see \S~\ref{S4.1}).  The ellipses in
  panels (e) and (f) are shown as every other to be visible at this
  size.  The red bar in panel (b) denotes a physical length of 2 kpc
  at the adopted distance.  }
\label{figcap10}
\end{figure}

\begin{figure}
\epsscale{1.0}
\plotone{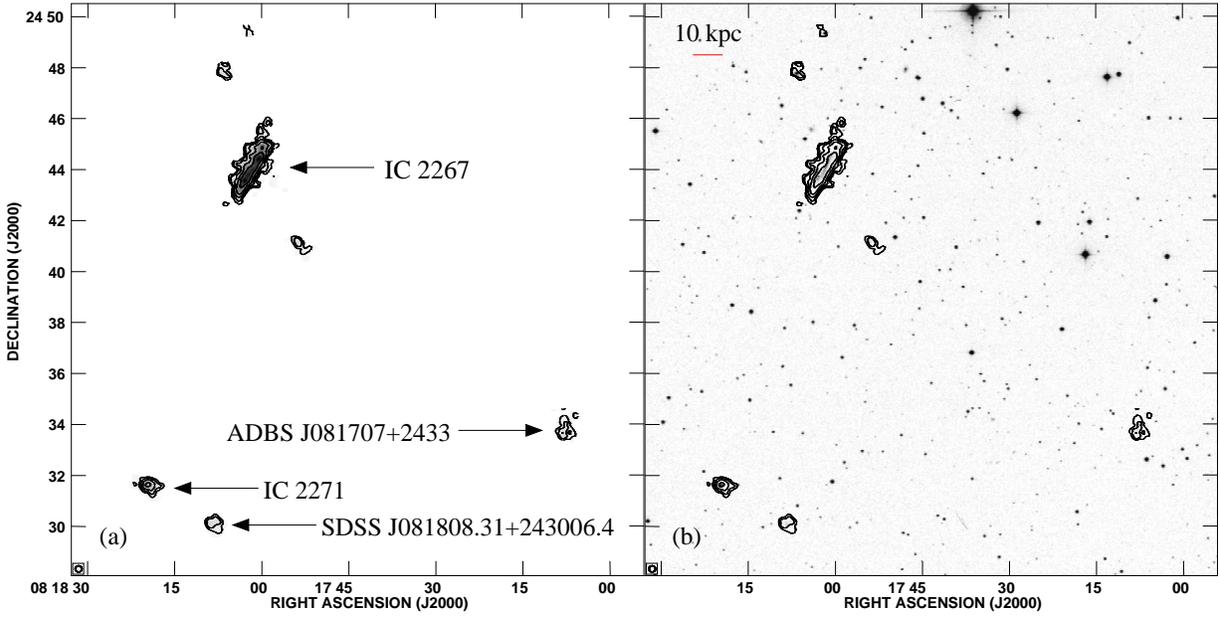}
\epsscale{1.0}
\caption{15\arcsec\ resolution \HI\ moment 0 (integrated
  \HI\ intensity) contours over \HI\ greyscale (a) and {\it Digitized
    Sky Survey} ({\it DSS}) r-band (b) images of the IC\,2271 loose
  association.The contours are at levels of (2, 4, 8, 16,
  32)\,$\times$\,10$^{20}$ cm$^{-2}$.  The minimum projected
  separations of IC\,2271 and its companions are as follows: 115 kpc
  for IC\,2267, 144 kpc for ADBS\,J081707+2433, and 26 kpc for SDSS
  J081808.31+243006.4.  The red bar in panel (b) denotes a physical
  length of 10 kpc at the adopted distance.}
\label{figcap11}
\end{figure}

\begin{figure}
\epsscale{1.0}
\plotone{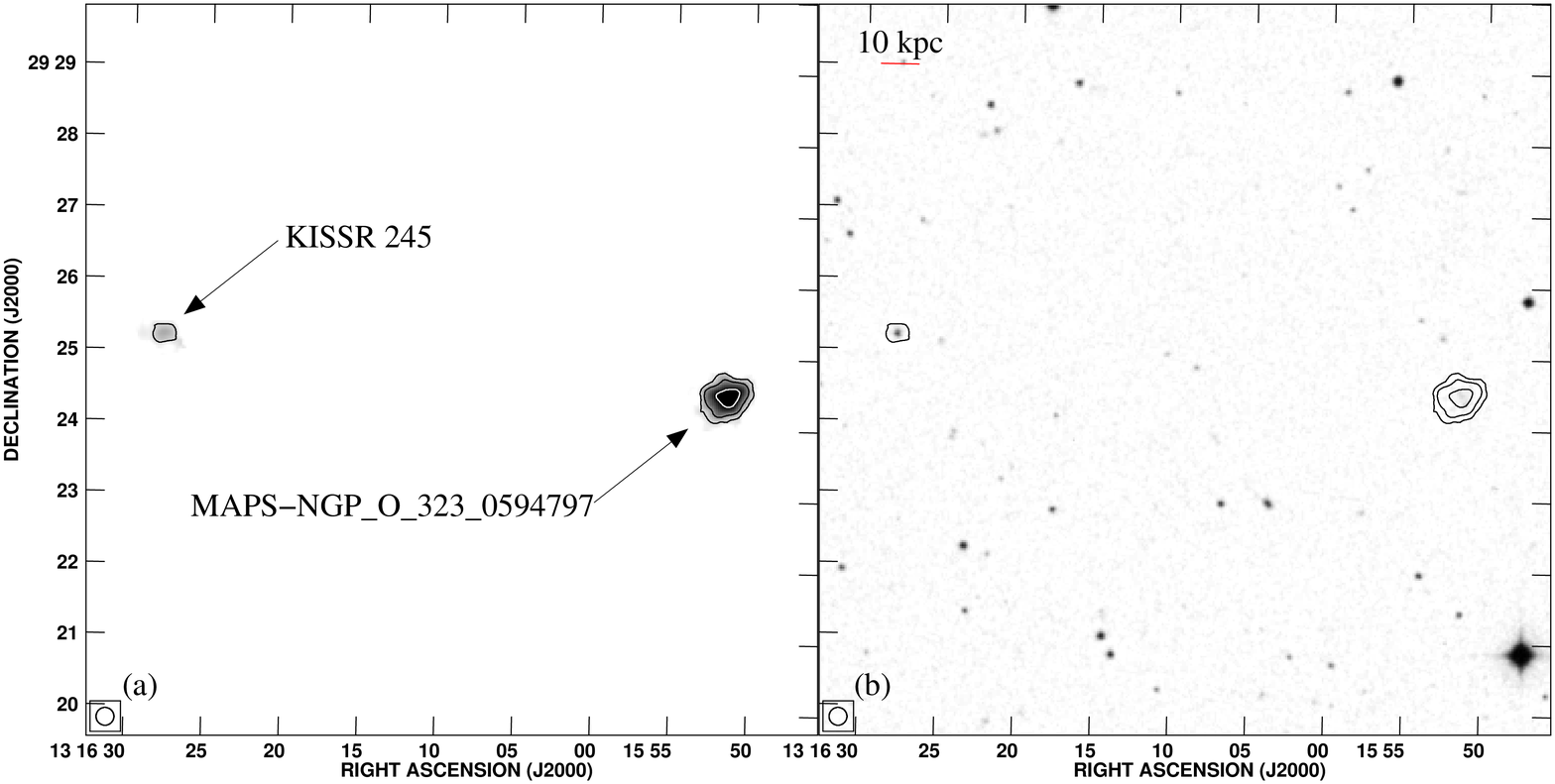}
\epsscale{1.0}
\caption{15\arcsec\ resolution \HI\ moment 0 (integrated
  \HI\ intensity) contours over \HI\ greyscale (a) and {\it DSS}
  r-band (b) images of the KISSR\,245 loose association. The contours
  are at levels of (1, 2, 4)\,$\times$\,10$^{20}$ cm$^{-2}$.  The
  minimum projected separation of KISSR\,245 and its companion
  (MAPS-NGP\_O\_323\_0594797) is 151 kpc.  The red bar in panel (b)
  denotes a physical length of 10 kpc at the adopted distance.}
\label{figcap12}
\end{figure}

\begin{figure}
\epsscale{1.0}
\plotone{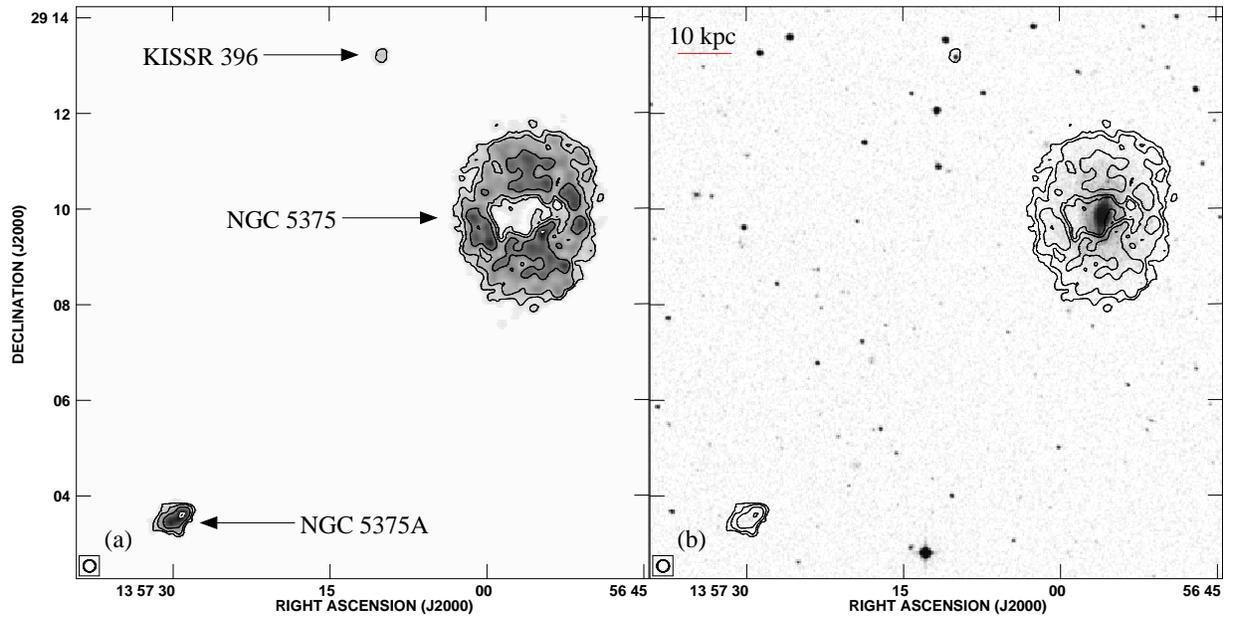}
\epsscale{1.0}
\caption{15\arcsec\ resolution \HI\ moment 0 (integrated
  \HI\ intensity) contours over \HI\ greyscale (a) and {\it DSS}
  r-band (b) images of the KISSR\,396 loose association. The contours
  are at levels of (1, 2, 4, 8)\,$\times$\,10$^{20}$ cm$^{-2}$.  The
  minimum projected separations of KISSR\,396 and its companions are
  as follows: 40 kpc for NGC\,5657 and 94 kpc for NGC\,5375A.  The red
  bar in panel (b) denotes a physical length of 10 kpc at the adopted
  distance.}
\label{figcap13}
\end{figure}

\begin{figure}
\epsscale{1.0}
\plotone{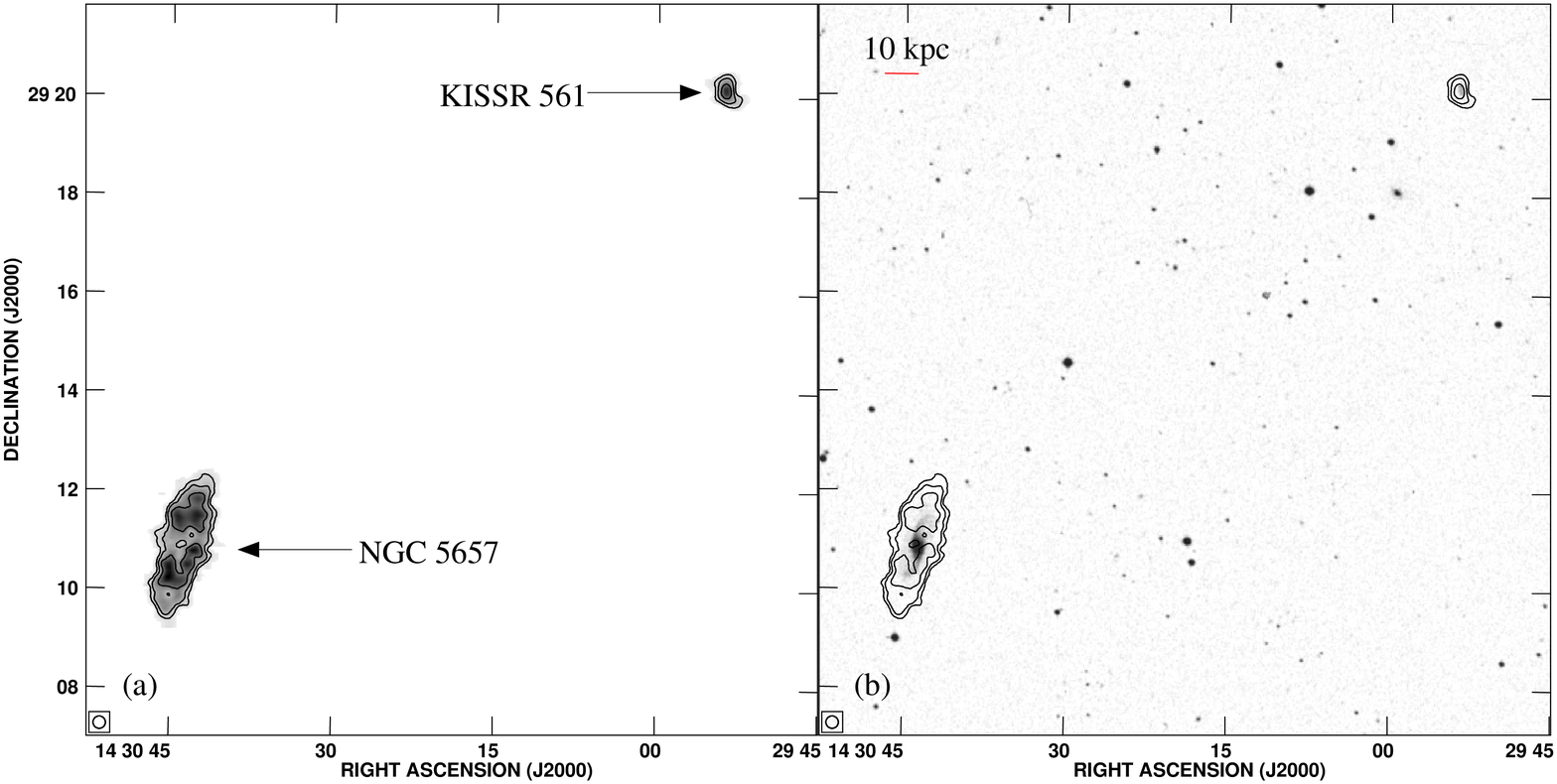}
\epsscale{1.0}
\caption{15\arcsec\ resolution \HI\ moment 0 (integrated
  \HI\ intensity) contours over \HI\ greyscale (a) and {\it DSS}
  r-band (b) images of the KISSR\,561 loose association. The contours
  are at levels of (2, 4, 8)\,$\times$\,10$^{20}$ cm$^{-2}$.  The
  minimum projected separation of KISSR\,561 and its companion
  (NGC\,5657) is 215 kpc.  The red bar in panel (b) denotes a physical
  length of 10 kpc at the adopted distance.}
\label{figcap14}
\end{figure}

\begin{figure}
\epsscale{1.0}
\plotone{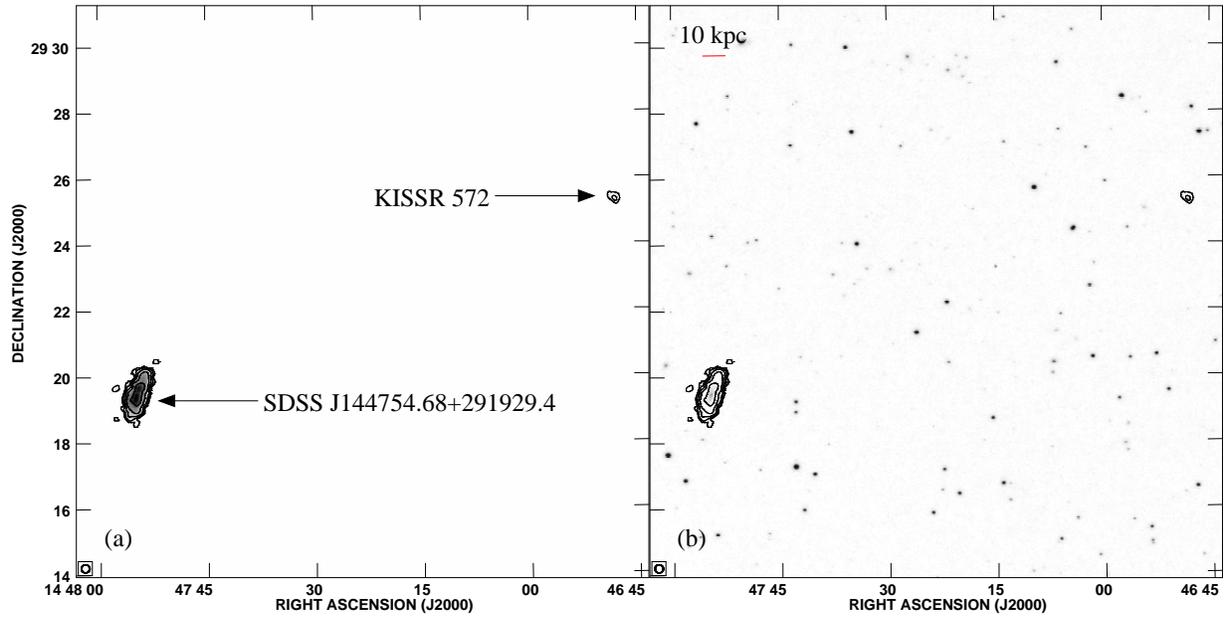}
\epsscale{1.0}
\caption{15\arcsec\ resolution \HI\ moment 0 (integrated
  \HI\ intensity) contours over \HI\ greyscale (a) and {\it DSS}
  r-band (b) images of the KISSR\,572 loose association. The contours
  are at levels of (0.6, 1.2, 2.4, 4.8, 9.6)\,$\times$\,10$^{20}$
  cm$^{-2}$.  The minimum projected separation of KISSR\,572 and its
  companion (SDSS\,J144754.68+291929.4) is 230 kpc.  The red bar in
  panel (b) denotes a physical length of 10 kpc at the adopted
  distance.}
\label{figcap15}
\end{figure}

\begin{figure}
\epsscale{1.0}
\plotone{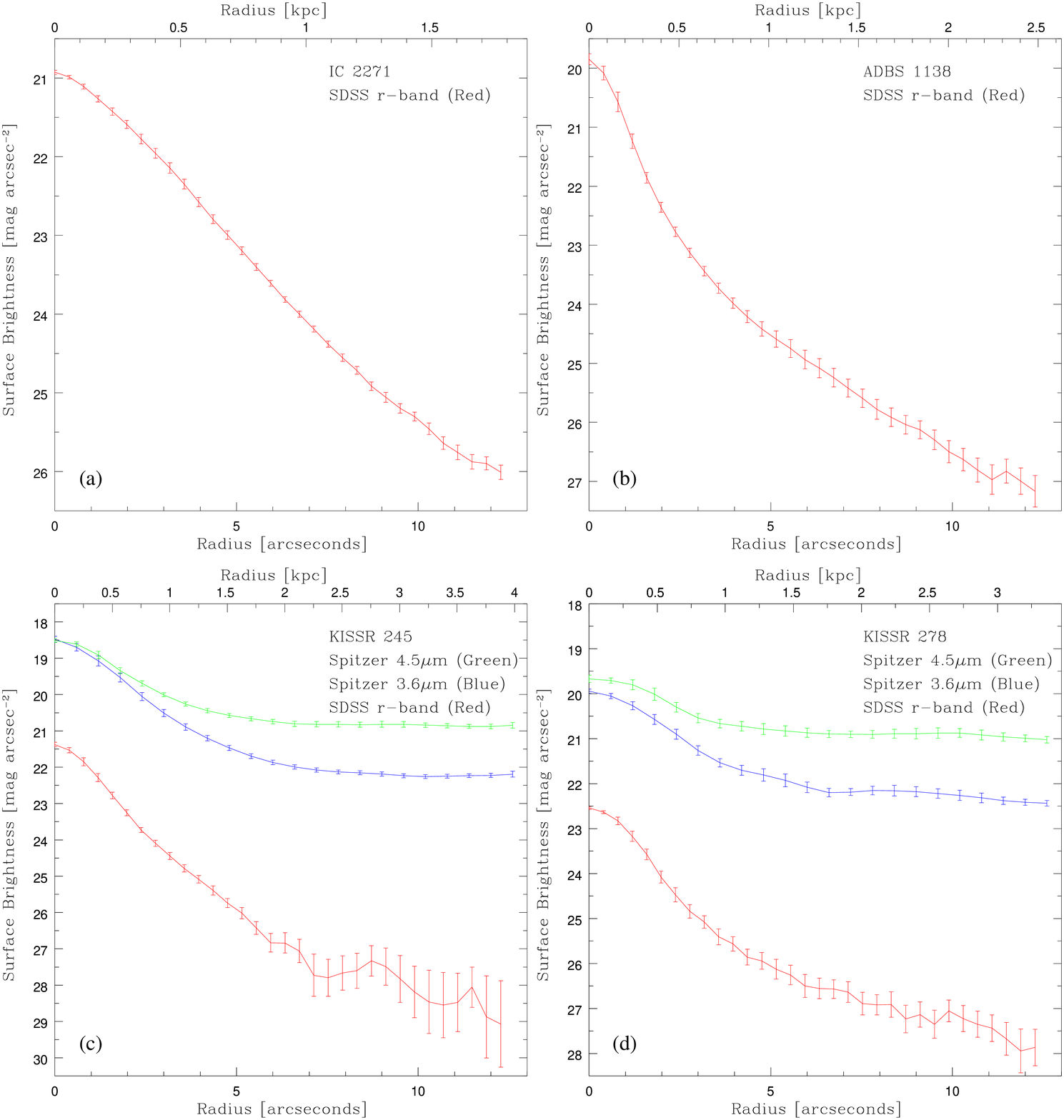}
\epsscale{1.0}
\caption{Optical and infrared surface brightness profiles of IC\,2271 (a),
   \adbs\ (b), KISSR\,245 (c), and KISSR\,278 (d).  Profiles are shown
  in the {\it SDSS} r-band (red), the {\it Spitzer} 3.6 $\mu$m band
  (blue), and the {\it Spitzer} 4.5 $\mu$m band (green).}
\label{figcap16}
\end{figure}

\begin{figure}
\epsscale{1.0}
\plotone{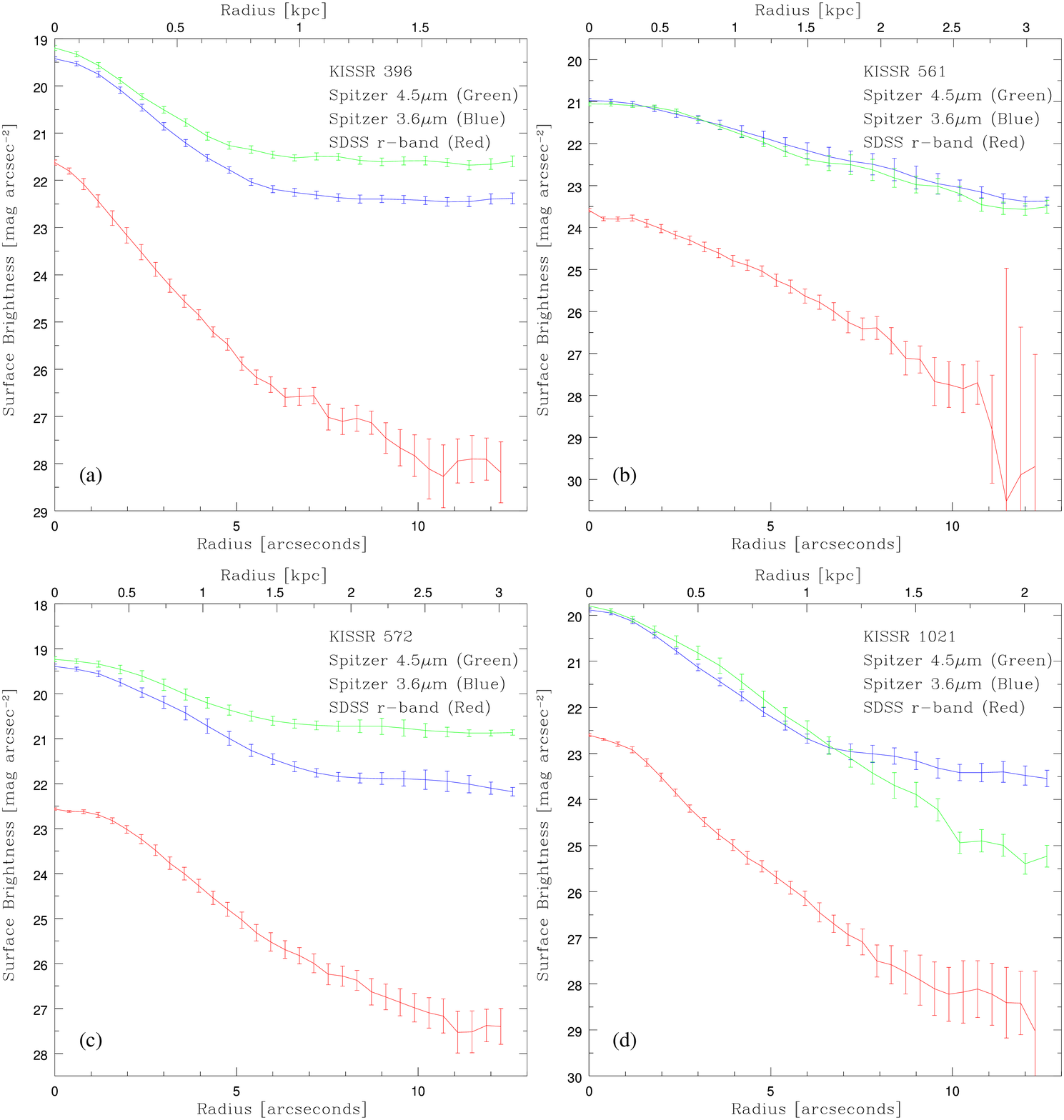}
\epsscale{1.0}
\caption{Same as Figure~\ref{figcap16}, for KISSR\,396 (a), KISSR\,561
  (b), KISSR\,572 (c), and KISSR\,1021 (d).}
\label{figcap17}
\end{figure}

\begin{figure}
\epsscale{1.0}
\plotone{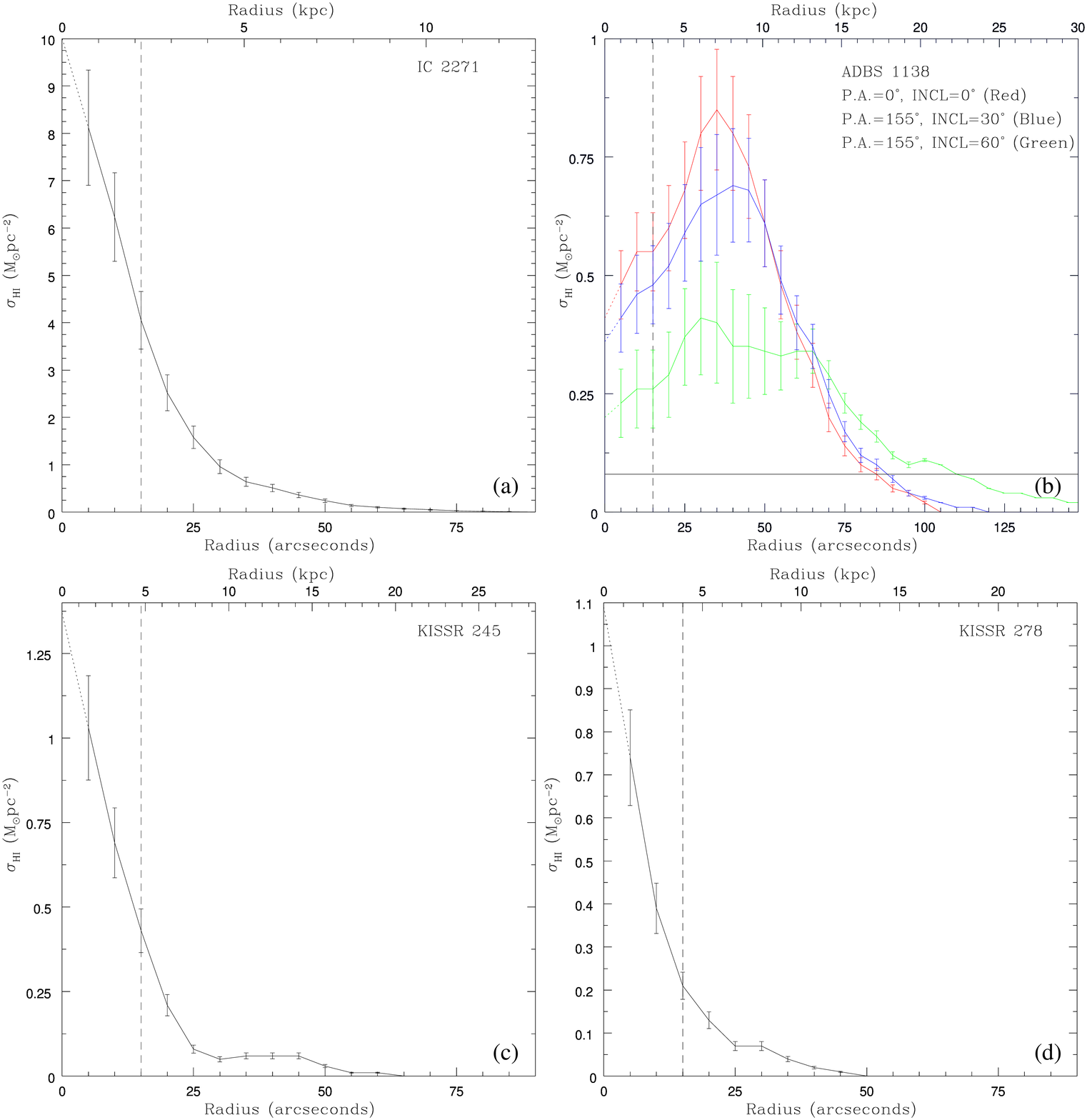}
\epsscale{1.0}
\caption{\HI\ radial column density profiles for IC\,2271 (a),
  \adbs\ (b), KISSR\,245 (c), and KISSR\,278 (d). These profiles were
  created by summing the \HI\ emission in concentric rings, with
  semi-major axis widths of 5\arcsec, from the center of each of the
  galaxies.  The vertical line denotes the \HI\ beam size.  The three
  profiles for \adbs\ show fits using the {\it SDSS} r-band
  P.A. (0$^{\circ}$) and Inclination (0$^{\circ}$) in Red, the
  P.A. (155$^{\circ}$) and Inclination (30$^{\circ}$) from
  \citet{cannon09} in Blue, and a higher Inclination (60$^{\circ}$)
  and P.A. (155$^{\circ}$) in Green. Note the \HI\ central depression
  found in the \adbs\ profile, and that at the point of 0.08
  \msun\ pc$^{-2}$ (denoted by a horizontal line) all of the profiles
  are roughly equal, only becoming slightly more extended with higher
  inclinations.}
\label{figcap18}
\end{figure}

\begin{figure}
\epsscale{1.0}
\plotone{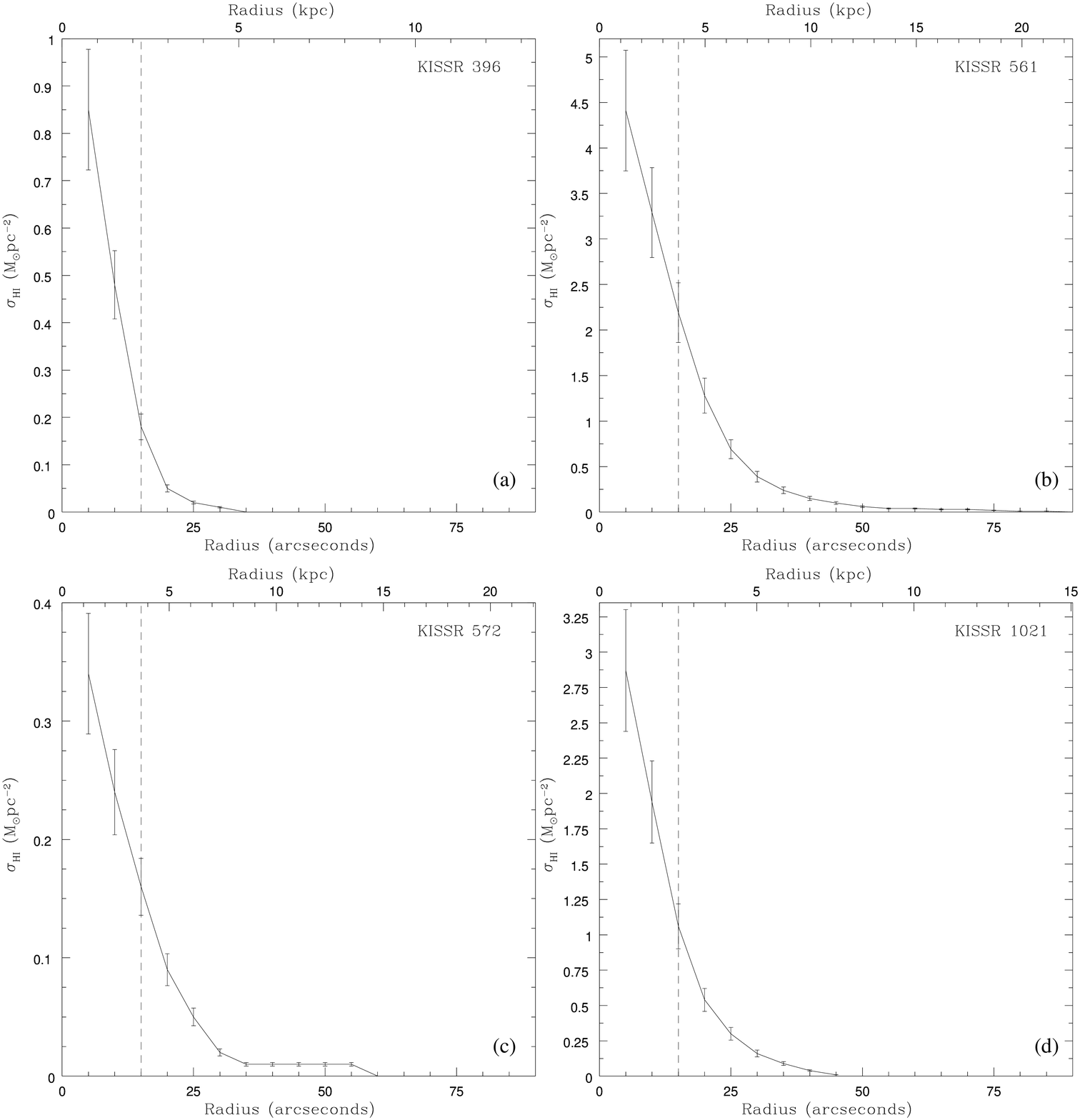}
\epsscale{1.0}
\caption{Same as Figure~\ref{figcap18}, for KISSR\,396 (a), KISSR\,561
  (b), KISSR\,572 (c), and KISSR\,1021 (d). These profiles were
  created by summing the \HI\ emission in concentric rings, with
  semi-major axis of 5\arcsec, from the center of each of the
  galaxies.}
\label{figcap19}
\end{figure}

We find that four of the eight systems (specifically, IC\,2271, \adbs,
KISSR\,561, and KISSR\,1021) have larger gaseous disks than stellar
disks; the remaining four (KISSR\,245, KISSR\,278, KISSR\,396, and
KISSR\,572) have only upper limits on the physical size of their
\HI\ disks (but note that KISSR\,245 has low-surface brightness
\HI\ gas extending well beyond a single beam size; see
Figure~\ref{figcap5}).  When parameterized by an \HI\ disk scale
length, the neutral gas disks of the seven available galaxies
(excluding \adbs) are larger than the stellar disks by factors of
$\sim$2--7.  When parameterized by the size of the \HI\ disk at the
10$^{19}$ cm$^{-2}$ column density level, the neutral gas disks of the
eight sample galaxies (including \adbs) are larger than the stellar
disks by factors of $\sim$4--58.  For each galaxy, the relative size
ratios measured using either \rhi\ or \hhi\ decrease with increasing
wavelength used to measure the stellar disk size (i.e., the optical
scale lengths are slightly shorter than the infrared scale lengths).
If the values of \rhi\ and \hhi\ are upper limits, then these ratios
decrease.

\adbs\ stands out among the members of this sample as harboring a
``giant'' \HI\ disk. While the \HI-to-optical size ratios of some
other members of the sample are substantial (e.g., IC\,2271,
KISSR\,245; see Table~\ref{t8}), this ratio for \adbs\ is more than
twice as large as any other measured here.  As discussed in
\citet{cannon09}, the physical properties of this galaxy are extreme
and enigmatic; \adbs\ has one of the largest known \HI-to-optical size
ratios in the local universe.  Given that the present sample was
selected {\it only} based on optically-derived properties, and that
the observations of the present sample were acquired and analyzed in a
uniform way, suggests that this ``giant disk'' dwarf galaxy is rare in
the local volume.  The present sample is small and is limited in
number by available (single dish \HI) data; the statistics behind the
present results could be significantly strengthened by a larger study
of additional extremely compact, star-forming dwarf galaxies in the
local volume.

\acknowledgements

This research was supported by a Cottrell College Science Award from
Research Corporation.  JLR acknowledges support from NSF
AST-000167932. The authors thank the National Radio Astronomy
Observatory and Macalester College for partial support of this work,
and Dr. Gustaaf van Moorsel for helpful discussions.  This work is
based in part on archival data obtained with the {\it Spitzer Space
  Telescope}, which is operated by the Jet Propulsion Laboratory,
California Institute of Technology under a contract with NASA. Support
for this work was provided by an award issued by JPL/Caltech.  This
work has made use of data from the {\it Sloan Digital Sky
  Survey}. Funding for the {\it SDSS} and {\it SDSS-II} has been
provided by the Alfred P. Sloan Foundation, the Participating
Institutions, the National Science Foundation, the U.S. Department of
Energy, the National Aeronautics and Space Administration, the
Japanese Monbukagakusho, the Max Planck Society, and the Higher
Education Funding Council for England. The {\it SDSS} Web Site is
http://www.sdss.org/.  The {\it SDSS} is managed by the Astrophysical
Research Consortium for the Participating Institutions. The
Participating Institutions are the American Museum of Natural History,
Astrophysical Institute Potsdam, University of Basel, University of
Cambridge, Case Western Reserve University, University of Chicago,
Drexel University, Fermilab, the Institute for Advanced Study, the
Japan Participation Group, Johns Hopkins University, the Joint
Institute for Nuclear Astrophysics, the Kavli Institute for Particle
Astrophysics and Cosmology, the Korean Scientist Group, the Chinese
Academy of Sciences (LAMOST), Los Alamos National Laboratory, the
Max-Planck-Institute for Astronomy (MPIA), the Max-Planck-Institute
for Astrophysics (MPA), New Mexico State University, Ohio State
University, University of Pittsburgh, University of Portsmouth,
Princeton University, the United States Naval Observatory, and the
University of Washington.  This research has made use of the NASA/IPAC
Extragalactic Database (NED) which is operated by the Jet Propulsion
Laboratory, California Institute of Technology, under contract with
the National Aeronautics and Space Administration, and NASA's
Astrophysics Data System.

\clearpage
\bibliographystyle{apj}                                                 


\clearpage
\begin{deluxetable}{lcccccc}
\tablecaption{Basic Parameters of Sample Galaxies} 
\tablewidth{0pt}  
\tablehead{ 
\colhead{Galaxy}  &\colhead{R.A.} &\colhead{DEC} &\colhead{V$_{\rm sys}$\tablenotemark{a}} &\colhead{Distance\tablenotemark{b}} &\colhead{\HI\ Mass} &\colhead{Single Dish \HI}\\
\colhead{Name}    &\colhead{(J2000)}  &\colhead{(J2000)}  &\colhead{(\kms)}     &\colhead{(Mpc)} &\colhead{(10$^7$ \msun)} &\colhead{Mass (10$^7$ \msun)}}
\startdata      
\vspace{0.0 cm}   
IC 2271    &08:18:19.7 &24:31:37 &2184 $\pm$ 10 &29.9 &25.1 $\pm$ 6.3 &101\tablenotemark{c,d}\\
ADBS 113845+2008 &11:38:45.2 &20:08:24 &3054 $\pm$ 3 &41.8 &40.2 $\pm$ 10.1 &84.6\tablenotemark{c} \\
KISSR\,245  &13:16:27.9 &29:25:11 &4762 $\pm$ 5 &65.2 &13.0 $\pm$ 3.3 &28.4\tablenotemark{e} \\
KISSR\,278  &13:23:37.7 &29:17:17 &4022 $\pm$ 5 &55.1 &4.4 $\pm$ 1.1 &12.0\tablenotemark{e} \\
KISSR\,396  &13:57:10.0 &29:13:10 &2238 $\pm$ 5 &30.7 &1.0 $\pm$ 0.3 &11.6\tablenotemark{e} \\
KISSR\,561  &14:29:53.6 &29:20:11 &3750 $\pm$ 5 &51.4 &34.3 $\pm$ 8.6 &56.5\tablenotemark{e} \\
KISSR\,572  &14:46:48.2 &29:25:18 &3693 $\pm$ 5 &50.6 &2.3 $\pm$ 0.6 &16.6\tablenotemark{e} \\
KISSR\,1021 &16:19:02.5 &29:10:22 &2519 $\pm$ 5 &34.5 &7.0 $\pm$ 1.8 &14.5\tablenotemark{e} \\
\enddata     
\label{t1}
\tablenotetext{a}{Derived from \HI\ global profiles using the full
  width at half maximum.}  \tablenotetext{b}{Assumes {\it H$_0$} = 73
  \kms\ Mpc$^{-1}$.}  \tablenotetext{c}{Calculated from the work of
  \citet{rosenberg00}.}  \tablenotetext{d}{\citet{rosenberg00} quote
  an {\it Arecibo} drift scan detection of IC\,2271 in the {\it ADBS}.
  Since single drift scan observations do not assure that a source
  passes through the center of the beam, the resulting flux integral
  is uncertain. Thus, the value quoted for IC\,2271 is from subsequent
  {\it VLA} D configuration snapshot observations.}
\tablenotetext{e}{Calculated from the work of \citet{lee02}.}
\end{deluxetable}

\clearpage
\begin{deluxetable}{lccccc}
\rotate
\tablecaption{{\it VLA} Observation Information} 
\tablewidth{0pt}  
\tablehead{ 
\colhead{Galaxy}  &\colhead{Date of} &\colhead{Total Time on} &\colhead{Total Bandwidth} &\colhead{Number of} &\colhead{Channel}\\
\colhead{Name}    &\colhead{Observation} &\colhead{Source (Minutes)} &\colhead{(MHz)} &\colhead{Channels} &\colhead{Separation (\kms)}}
\startdata      
\vspace{0.0 cm}      
IC\,2271  &2006 Oct. 31 &241 &3.12 &64 &10.30\\
ADBS\,113845+2008  &2006 Oct. 31 &245 &1.56 &128 &2.58\\
KISSR\,245  &2008 Apr. 23 &252 &3.12 &128 &5.15\\
KISSR\,278  &2008 Apr. 29 &253 &3.12 &128 &5.15\\
KISSR\,396  &2008 May 1 &254 &3.12 &128 &5.15\\
KISSR\,561  &2008 May 13 &252 &3.12 &128 &5.15\\
KISSR\,572  &2008 May 8 &251 &3.12 &128 &5.15\\
KISSR\,1021  &2008 May 15 &255 &3.12 &128 &5.15\\
\enddata     
\label{t2}
\end{deluxetable} 

\clearpage
\begin{deluxetable}{lcccc}
\tablecaption{{\it VLA} Observations of the Compact Dwarfs Sample} 
\tablewidth{0pt}  
\tablehead{ 
\colhead{Galaxy}   &\colhead{Velocity} &\colhead{RMS noise}  &\colhead{S$_{\rm HI}$ ({\it VLA})}  &\colhead{S$_{\rm HI}$ (Single Dish)}\\
\colhead{Name}     &\colhead{Range (\kms)}     &\colhead{(mJy Bm$^{-1}$)} &\colhead{(Jy \kms)} &\colhead{(Jy \kms)}}
\startdata      
\vspace{0.0 cm}
IC\,2271     &2121 - 2246 &0.39 &1.22 $\pm$ 0.12 &4.702\tablenotemark{a,b}\\
ADBS\,113845+2008  &2987 - 3117 &0.73 &0.98 $\pm$ 0.10 &1.982\tablenotemark{a} \\
KISSR\,245   &4744 - 4779 &0.52 &0.13 $\pm$ 0.01 &0.28\tablenotemark{c} \\
KISSR\,278   &4014 - 4030 &0.48 &0.062 $\pm$ 0.006 &0.17\tablenotemark{c} \\
KISSR\,396   &2221 - 2253 &0.53 &0.043 $\pm$ 0.004 &0.51\tablenotemark{c} \\
KISSR\,561   &3695 - 3809 &0.50 &0.55 $\pm$ 0.06 &0.89\tablenotemark{c} \\
KISSR\,572   &3688 - 3699 &0.48 &0.038 $\pm$ 0.004 &0.27\tablenotemark{c} \\
KISSR\,1021  &2473 - 2562 &0.46 &0.25 $\pm$ 0.03 &0.51\tablenotemark{c} \\
\enddata     
\label{t3}
\tablenotetext{a}{\citet{rosenberg00}} 
\tablenotetext{b}{\citet{rosenberg00} quote an {\it Arecibo} drift scan detection of IC\,2271 in the {\it ADBS}.  Since single drift scan observations do not assure that a source passes through the center of the beam, the resulting flux integral is uncertain. Thus, the value quoted for IC\,2271 is from subsequent {\it VLA} D configuration snapshot observations.}
\tablenotetext{c}{\citet{lee02}}
\end{deluxetable}  

\clearpage
\begin{deluxetable}{lccc}
\tablecaption{UVTAPER Tests} 
\tablewidth{0pt}  
\tablehead{ 
\colhead{Galaxy} &\colhead{} &\colhead{S$_{\rm HI}$ (Jy \kms)} &\colhead{}\\
\colhead{Name}   &\colhead{15\arcsec\tablenotemark{a}} &\colhead{30\arcsec\tablenotemark{a}} &\colhead{60\arcsec\tablenotemark{a}}}
\startdata    
IC\,2271      &1.2 &1.5 &1.5 \\
ADBS\,113845+2008   &0.98 &0.90 &0.90 \\
KISSR\,245    &0.13 &0.14 &0.15 \\
KISSR\,278    &0.062 &0.30 &-\tablenotemark{b} \\
KISSR\,396    &0.043 &0.038 &0.039 \\
KISSR\,561    &0.55 &0.65 &0.66 \\
KISSR\,572    &0.038 &0.028 &-\tablenotemark{b} \\
KISSR\,1021   &0.25 &0.35 &0.23 \\
\enddata     
\label{t4}
\tablenotetext{a}{Circular beamsize created from changing the
  parameter UVTAPER in the AIPS task IMAGR.}  
\tablenotetext{b}{Galaxy could not be found in three or more 
  consecutive channels when blanking through the cube.}
\end{deluxetable}  

\clearpage
\begin{deluxetable}{lcccc}
\rotate
\tablecaption{Relative Distances Between Companion Galaxies} 
\tablewidth{0pt}  
\tablehead{ 
\colhead{Target}   &\colhead{Companion} &\colhead{Positional Offset}  &\colhead{Velocity Offset} &\colhead{Minimum Separation}\\
\colhead{Galaxy}   &\colhead{Galaxy(s)} &\colhead{(arcsec)}           &\colhead{(\kms)}          &\colhead{(kpc)}}
\startdata      
\vspace{0.0 cm}     
IC\,2271     &ADBS J081707+2433         &995  &121  &144 \\
             &IC 2267                   &790  &146  &115 \\
             &SDSS J081808.31+243006.4  &180  &105  &26 \\
\hline
KISSR\,245   &MAPS-NGP\_O\_323\_0594797 &477  &12   &151 \\
\hline
KISSR\,396   &NGC 5375                  &270  &137  &40 \\
             &NGC 5375A                 &633  &18   &94 \\
\hline
KISSR\,561   &NGC 5657                  &861  &110  &215 \\
\hline
KISSR\,572   &SDSS J144754.68+291929.4  &936  &33   &230 \\
\enddata     
\label{t5}
\end{deluxetable}  

\clearpage
\begin{deluxetable}{lcc}
\tablecaption{Eillipse Parameters for Isophotal Photometry} 
\tablewidth{0pt}  
\tablehead{ 
\colhead{Galaxy} &\colhead{P.A.}      &\colhead{Inclination}  \\
\colhead{Name}   &\colhead{E of N}  &\colhead{} }
\startdata      
\vspace{0.0 cm}
IC\,2271     &143$^{\circ}$ &45$^{\circ}$ \\
ADBS\,113845+2008  &0$^{\circ}$ &0$^{\circ}$ \\
KISSR\,245   &0$^{\circ}$ &0$^{\circ}$ \\
KISSR\,278   &40$^{\circ}$ &73$^{\circ}$ \\
KISSR\,396   &175$^{\circ}$ &43$^{\circ}$ \\
KISSR\,561   &0$^{\circ}$ &57$^{\circ}$ \\
KISSR\,572   &60$^{\circ}$ &70$^{\circ}$ \\
KISSR\,1021  &130$^{\circ}$ &48$^{\circ}$ \\
\enddata     
\label{t6}
\end{deluxetable}  

\clearpage
\begin{deluxetable}{lccccc}
\tablecaption{Scale Length Calculations} 
\tablewidth{0pt}  
\tablehead{ 
\colhead{Galaxy} &\colhead{\hr}    &\colhead{\hthree} &\colhead{\hfour} &\colhead{\hhi}   &\colhead{\rhi}\\
\colhead{Name}   &\colhead{(kpc)}  &\colhead{(kpc)}  &\colhead{(kpc)}  &\colhead{(kpc)}   &\colhead{(kpc)}}
\startdata      
\vspace{0.0 cm}     
IC\,2271             &0.33 $\pm$ 0.08 &N/A             &N/A             &1.72 $\pm$ 0.43    &8.74 $\pm$ 2.19\\
ADBS\,113845+2008    &0.30 $\pm$ 0.08 &N/A             &N/A             &-\tablenotemark{a} &17.3 $\pm$ 4.3 \\
KISSR\,245           &0.38 $\pm$ 0.10 &0.55 $\pm$ 0.14 &0.70 $\pm$ 0.18 &2.48 $\pm$ 0.62    &$<$8.28 $\pm$ 2.07\\
KISSR\,278           &0.41 $\pm$ 0.10 &0.80 $\pm$ 0.20 &1.36 $\pm$ 0.34 &2.64 $\pm$ 0.66    &$<$6.19 $\pm$ 1.55\\
KISSR\,396           &0.19 $\pm$ 0.05 &0.31 $\pm$ 0.08 &0.36 $\pm$ 0.09 &0.78 $\pm$ 0.20    &$<$2.65 $\pm$ 0.66\\
KISSR\,561           &0.72 $\pm$ 0.18 &1.30 $\pm$ 0.33 &1.20 $\pm$ 0.30 &2.59 $\pm$ 0.65    &11.8 $\pm$ 3.0 \\
KISSR\,572           &0.50 $\pm$ 0.13 &0.77 $\pm$ 0.19 &1.18 $\pm$ 0.30 &2.19 $\pm$ 0.55    &$<$4.64 $\pm$ 1.16\\
KISSR\,1021          &0.29 $\pm$ 0.07 &0.37 $\pm$ 0.09 &0.31 $\pm$ 0.08 &1.18 $\pm$ 0.30    &5.39 $\pm$ 1.35\\
\hline                                                                                                                
\hline                                                                                                                
\colhead{Galaxy} &\colhead{\hr}      &\colhead{\hthree}  &\colhead{\hfour}   &\colhead{\hhi}      &\colhead{\rhi}\\   
\colhead{Name}   &\colhead{(arcsec)} &\colhead{(arcsec)} &\colhead{(arcsec)} &\colhead{(arcsec)}  &\colhead{(arcsec)}\
\\                                                                                                                    
\hline                                                                                                                
IC\,2271             &2.29 $\pm$ 0.57 &N/A             &N/A             &11.9 $\pm$ 3.0     &60.3 $\pm$ 15.1 \\       
ADBS\,113845+2008    &1.44 $\pm$ 0.36 &N/A             &N/A             &-\tablenotemark{a} &85.3 $\pm$ 21.3 \\       
KISSR\,245           &1.18 $\pm$ 0.30 &1.71 $\pm$ 0.43 &2.21 $\pm$ 0.55 &7.84 $\pm$ 1.96    &$<$26.2 $\pm$ 6.6 \\        
KISSR\,278           &1.51 $\pm$ 0.38 &2.99 $\pm$ 0.75 &5.12 $\pm$ 1.28 &9.88 $\pm$ 2.47    &$<$23.2 $\pm$ 5.8 \\        
KISSR\,396           &1.33 $\pm$ 0.33 &2.13 $\pm$ 0.53 &2.50 $\pm$ 0.63 &5.25 $\pm$ 1.31    &$<$17.8 $\pm$ 4.5 \\        
KISSR\,561           &2.93 $\pm$ 0.73 &5.27 $\pm$ 1.32 &4.86 $\pm$ 1.22 &10.4 $\pm$ 2.6     &47.4 $\pm$ 11.9 \\       
KISSR\,572           &2.07 $\pm$ 0.52 &3.18 $\pm$ 0.80 &4.87 $\pm$ 1.22 &8.93 $\pm$ 2.23    &$<$18.9 $\pm$ 4.7 \\        
KISSR\,1021          &1.74 $\pm$ 0.44 &2.23 $\pm$ 0.56 &1.86 $\pm$ 0.47 &7.05 $\pm$ 1.76    &32.2 $\pm$ 8.1 \\        
\enddata     
\label{t7}
\tablenotetext{a}{The value of \hhi\ for \adbs\ cannot be accurately
  determined due to the \HI\ central depression and outer high column
  density ring.}
\end{deluxetable}

%
%

\clearpage
\begin{deluxetable}{lcccccc}
\tablecaption{Relative Sizes of the Stellar and Gaseous Disks} 
\tablewidth{0pt}  
\tablehead{ 
\colhead{Galaxy}   &\colhead{\hhi/\hr} &\colhead{\rhi/\hr}  &\colhead{\hhi/\hthree} &\colhead{\rhi/\hthree}  &\colhead{\hhi/\hfour} &\colhead{\rhi/\hfour}}
\startdata      
\vspace{0.0 cm}     
IC\,2271            &5.2  $\pm$ 1.8      &26 $\pm$ 9       &N/A           &N/A            &N/A           &N/A \\
ADBS\,113845+2008   &-\tablenotemark{a}  &58 $\pm$ 21      &N/A           &N/A            &N/A           &N/A \\
KISSR\,245          &6.5  $\pm$ 2.4      &$<$22 $\pm$ 8    &4.5 $\pm$ 1.6 &$<$15 $\pm$ 5     &3.5 $\pm$ 1.3 &$<$12 $\pm$ 4.2 \\
KISSR\,278          &6.4  $\pm$ 2.2      &$<$15 $\pm$ 5    &3.3 $\pm$ 1.2 &$<$7.7 $\pm$ 2.7  &1.9 $\pm$ 0.7 &$<$4.6 $\pm$ 1.6 \\
KISSR\,396          &4.1  $\pm$ 1.5      &$<$14 $\pm$ 5    &2.5 $\pm$ 0.9 &$<$8.5 $\pm$ 3.0  &2.2 $\pm$ 0.8 &$<$7.4 $\pm$ 2.6 \\
KISSR\,561          &3.6  $\pm$ 1.3      &16 $\pm$ 6       &2.0 $\pm$ 0.7 &9.1 $\pm$ 3.2  &2.2 $\pm$ 0.8 &9.8 $\pm$ 3.5 \\
KISSR\,572          &4.4  $\pm$ 1.6      &$<$9.3 $\pm$ 3.3 &2.8 $\pm$ 1.0 &$<$6.0 $\pm$ 2.1  &1.9 $\pm$ 0.7 &$<$3.9 $\pm$ 1.4 \\
KISSR\,1021         &4.1  $\pm$ 1.4      &19 $\pm$  6      &3.2 $\pm$ 1.1 &14.6 $\pm$ 5.2 &3.8 $\pm$ 1.4 &17 $\pm$ 6.2 \\
\enddata     
\label{t8}
\tablenotetext{a}{The value of \hhi\ for \adbs\ cannot be accurately
  determined due to the \HI\ central depression and outer high column
  density ring.}
\end{deluxetable}  

\clearpage

\clearpage

\clearpage

\clearpage

\clearpage

\clearpage

\clearpage

\clearpage

\clearpage

\clearpage

\clearpage

\clearpage

\clearpage

\clearpage

\clearpage

\clearpage

\clearpage

\clearpage

\clearpage

\end{document}